\documentclass[english,11pt,oneside]{article}

\pdfoutput=1
\usepackage[T1]{fontenc}
\usepackage[latin9]{inputenc}
\usepackage[usenames,dvipsnames]{xcolor}
\usepackage{color}
\usepackage{babel}
\usepackage{setspace}
\usepackage{amstext}
\usepackage{amssymb}
\PassOptionsToPackage{normalem}{ulem}
\usepackage{ulem}
\usepackage[unicode=true,pdfusetitle,
 bookmarks=true,bookmarksnumbered=false,bookmarksopen=false,citecolor=Turquoise,
 breaklinks=false,pdfborder={0 0 1},backref=false,colorlinks=true,pdfpagemode=FullScreen,linktoc=page]
 {hyperref}
\usepackage{graphicx}
\usepackage{mathtools}
\usepackage[small]{caption}
\usepackage{changepage}
\usepackage{slashed}
\usepackage{slashbox}
\usepackage{makecell}
\usepackage{multirow}
\providecommand{\tabularnewline}{\\}
\definecolor{note_fontcolor}{rgb}{0.80078125, 0.80078125, 0.80078125}
\usepackage[top=3 cm, bottom=3 cm, left=3.1416 cm, right=3.1416 cm]{geometry}

\def\beq{\begin{equation}}
\def\eeq{\end{equation}}
\def\bea{\begin{eqnarray}}
\def\eea{\end{eqnarray}}

\begin{document}

\baselineskip=17pt

%%%%%%%%%%
%%%%%%%%%%    Title page
%%%%%%%%%%

\thispagestyle{empty}
\vspace{20pt}
\font\cmss=cmss10 \font\cmsss=cmss10 at 7pt

\begin{flushright}
%\today \\
%
UMD-PP-016-006\\
\end{flushright}

\hfill
%\vspace{20pt}

\begin{center}
{\Large \textbf
{
Flavor Universal Resonances and Warped Gravity
%
%Flavor Universal Resonances from Warped/Composite Framework
%
}}
\end{center}

\vspace{15pt}

\begin{center}
{\large Kaustubh Agashe$\, ^{a}$, Peizhi Du$\, ^{a}$, Sungwoo Hong$\, ^{a}$, Raman Sundrum$\, ^{a}$ \\
\vspace{15pt}
$^{a}$\textit{Maryland Center for Fundamental Physics,
     Department of Physics,
     University of Maryland,
     College Park, MD 20742, U.~S.~A.} \\
   
\vspace{0.3cm}
      
{\it email addresses}: kagashe@umd.edu, pdu@umd.edu, sungwoo83hong@gmail.com, raman@umd.edu}

\end{center}

\vspace{5pt}

\begin{center}
\textbf{Abstract}
\end{center}
\vspace{5pt} 

Warped higher-dimensional compactifications with ``bulk'' standard model, or their AdS/CFT dual as the purely 4D scenario of Higgs compositeness and partial compositeness, offer an elegant approach to resolving the electroweak hierarchy problem as well as the origins of flavor structure.
However, low-energy electroweak/flavor/CP constraints and the absence of non-standard physics at LHC Run 1 suggest that a ``little hierarchy problem'' remains, and that the new physics underlying naturalness may lie 
%
%modestly 
%
out of LHC reach.
Assuming this to be the case, we show that there is a simple and natural extension of the minimal warped model in the Randall-Sundrum framework, in which matter, gauge and gravitational fields propagate modestly different degrees into the IR of the warped dimension, resulting in rich and striking consequences for the LHC (and beyond). The LHC-accessible part of the new physics is AdS/CFT dual to the mechanism of ``vectorlike confinement'', with TeV-scale Kaluza-Klein excitations of the gauge and gravitational fields dual to spin-0,1,2 composites. Unlike the minimal warped model, 
these 
low-lying excitations have predominantly flavor-blind and flavor/CP-safe interactions with the standard model.
%
%mediated by standard gauge fields. 
%
Remarkably, this scenario also predicts small deviations from flavor-blindness originating from virtual effects of Higgs/top compositeness at $\sim O(10)$ TeV, with subdominant resonance decays into Higgs/top-rich final states, giving the LHC an early ``preview'' of the nature of the resolution of the hierarchy problem. Discoveries of this type at LHC Run 2 would thereby anticipate (and set a target for) even more explicit explorations of Higgs compositeness at a 100 TeV collider, or for next-generation flavor tests.

\vfill\eject
\noindent

%%%%%%%%%%
%%%%%%%%%%    Main Text
%%%%%%%%%%

\section{Introduction}
\label{sec: Introduction}

The scenario of Higgs compositeness \cite{Panico:2015jxa} offers a powerful resolution to the Hierarchy Problem. The Standard Model (SM) Higgs degrees of freedom remain much lighter than the Planck scale in the face of radiative corrections because they are only assembled at $\sim$ TeV scale, as tightly bound 
composites of some new strongly interacting ``preons''. This is in close analogy to how the ordinary charged pion remains much lighter than the Planck scale in the face of QED radiative corrections, by being assembled as a quark-gluon composite at $\sim$ GeV. But despite the simple plot, composite Higgs dynamics is notoriously difficult to model in detail because it requires understanding a new strongly-coupled dynamics, operating outside perturbative control. 

\begin{figure}
\center
\includegraphics[width=0.7\linewidth]{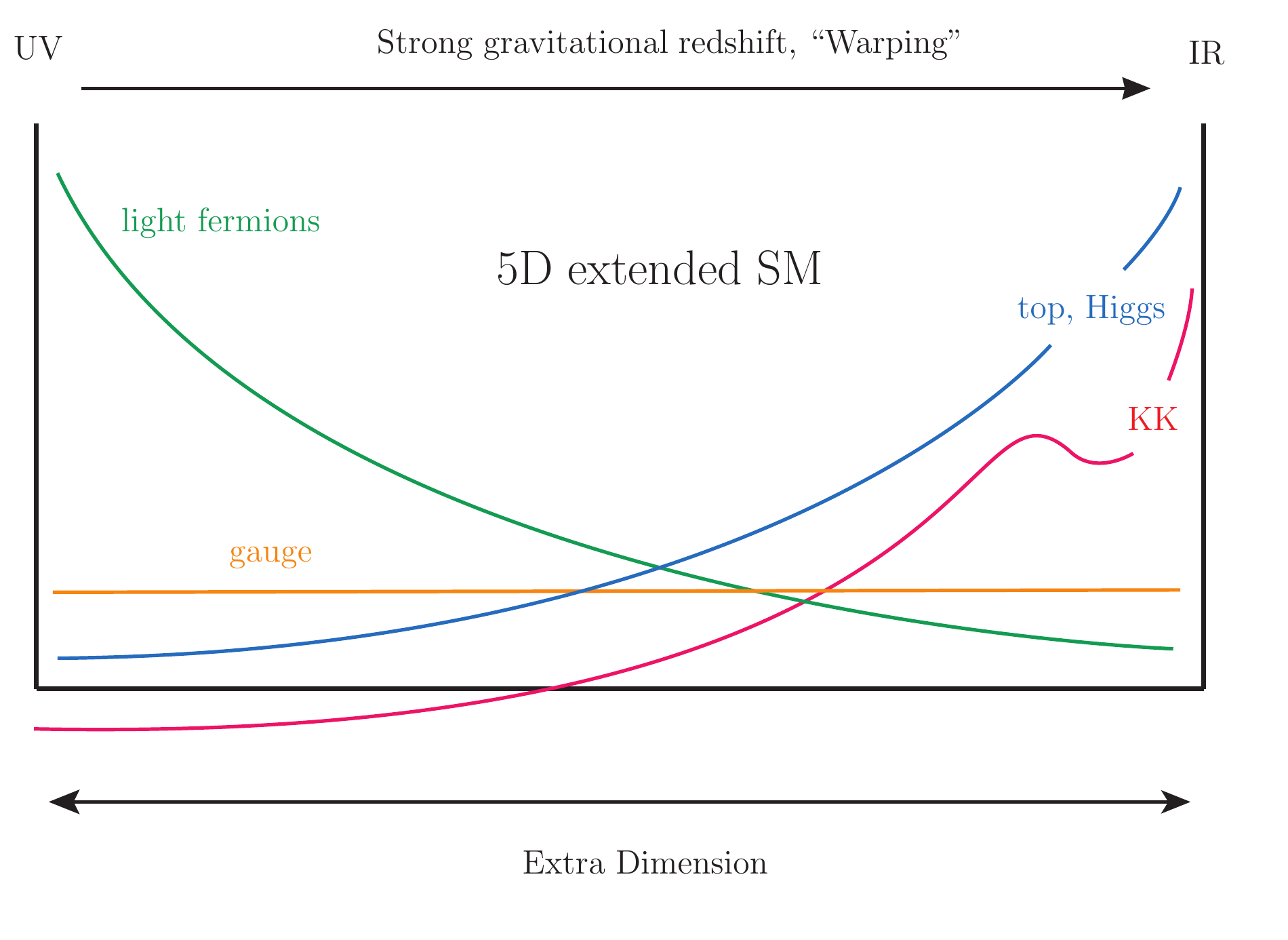}

\caption{Minimal RS1 model with SM fields in bulk. Also shown are schematic shapes of extra-dimensional wavefunctions for various particles (zero modes/SM and a generic KK mode).}
\label{fig:OriginalRS}
\end{figure}

Remarkably, Higgs compositeness has an alternate ``dual'' formulation \cite{ArkaniHamed:2000ds, Rattazzi:2000hs, Contino:2004vy, Contino:2003ve, Agashe:2004rs} in the form of ``warped'' higher-dimensional theories of Randall-Sundrum type \cite{Randall:1999ee}, related to the purely 4D formulation via the famous AdS/CFT correspondence \cite{Aharony:1999ti}. In the warped framework there can exist a regime of weakly-coupled higher-dimensional effective field theory, allowing more detailed phenomenological modeling as well as a prototype for UV completion, say within string theory \cite{Strassler:2003ht}.  Fig.~\ref{fig:OriginalRS} shows a schematic representation of particle physics in the simplest such setting, with a single microscopic extra-dimensional interval. The SM is now fundamentally 5-dimensional \cite{Gherghetta:2006ha}, but its lightest modes appear as the familiar 4D SM particles, with phenomenological properties deriving from their extra-dimensional wavefunctions. In particular, the SM fermions naturally have disparate  wavefunctions, which lead to an attractive mechanism for the origin of SM flavor structure, AdS/CFT dual to the robust mechanism of Partial Compositeness \cite{Kaplan:1991dc}.

\begin{figure}
\center
\includegraphics[width=0.4\linewidth]{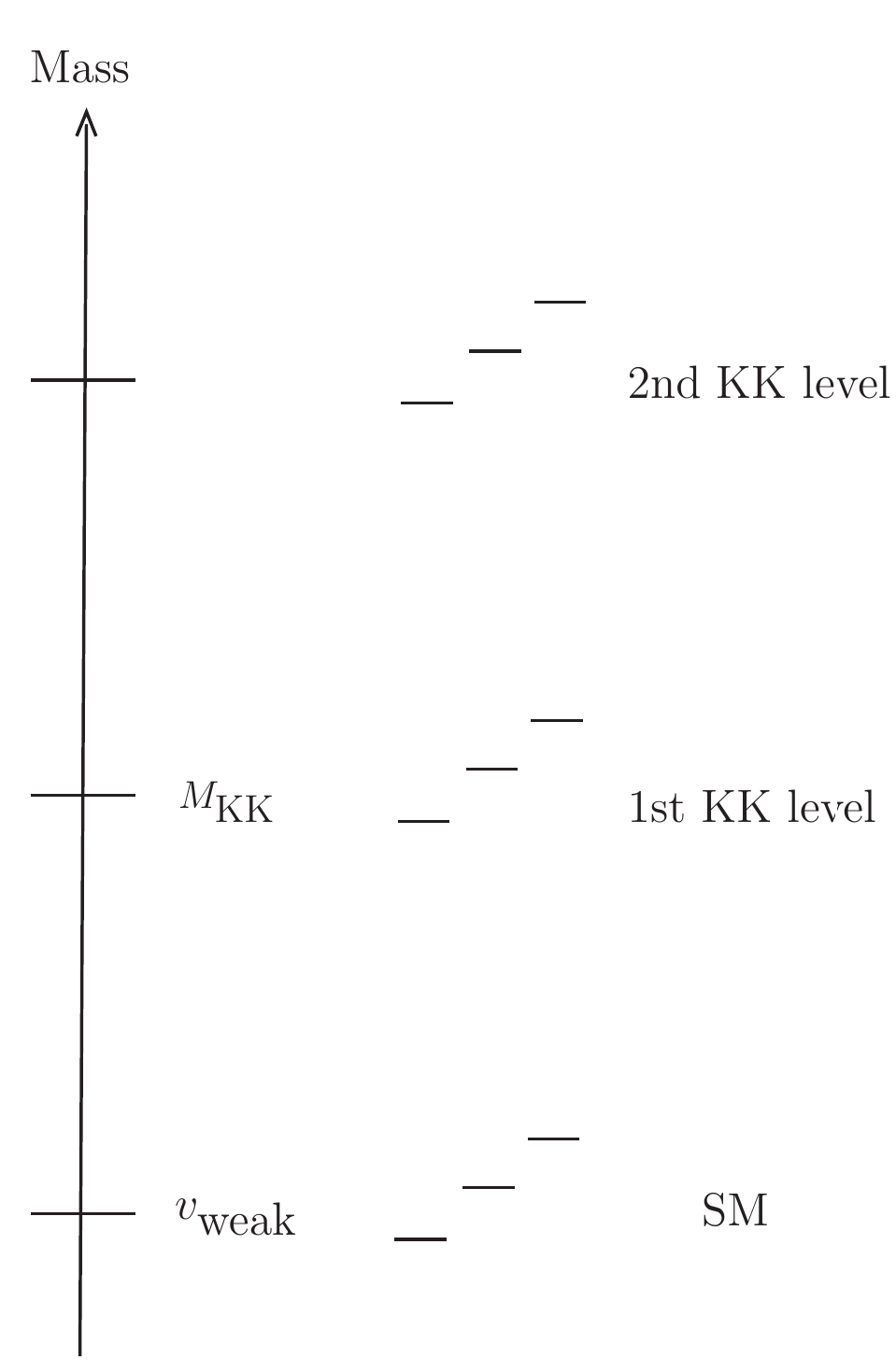}

\caption{General spectrum of model of Fig.~\ref{fig:OriginalRS}.}
\label{fig:Generalspectrum}
\end{figure}

On top of the lightest modes are Kaluza-Klein (KK) excitations of the SM (Fig.~\ref{fig:Generalspectrum}), which effectively cut off quantum corrections to the Higgs mass and electroweak symmetry breaking (EWSB). Naturalness then implies that these KK states should have masses of the order TeV scale.\footnote{An elegant realization in warped extra dimension of the composite Higgs mechanism, i.e., where it is a PNGB like the pion, is via gauge-Higgs unification \cite{Contino:2003ve, Agashe:2004rs}. It is in 
this case that the cutoff of Higgs quantum corrections is the KK scale.
However, this aspect plays little role in this paper. 
So, for brevity, we simply suppress this extra structure of the Higgs field. %we simply treat the Higgs boson as a scalar from the 5D viewpoint, with a profile sharply peaked near the IR brane.
}
This is the basis of ongoing LHC searches for KK-excited  tops and bottoms (``top partners'') and KK gauge bosons and spin-2 KK gravitons.
Because of their strong extra-dimensional wavefunction-overlap with the top quark and Higgs, these KK resonances predominantly decay to $t, h, W_L, Z_L$ \cite{Agashe:2013kyb}. From the viewpoint of 4D Higgs compositeness, the KK excitations are simply other composites of the same preons inside the Higgs (and the closely-related top quark).

Lower-energy experiments are also sensitive to KK states via their {\it virtual} exchanges. Electroweak precision tests, now including the rapidly developing body of precision Higgs measurements, robustly constrain the KK spectrum, but are still consistent with 
KK discoverability at the LHC \cite{Agashe:2003zs}. However, as in the supersymmetric paradigm, the constraints from tests of flavor and CP violation are extremely 
stringent. Although the warped extra-dimensional framework (and partial compositeness) enjoys a powerful generalization of the SM GIM mechanism suppressing FCNCs \cite{Gherghetta:2000qt}, it is imperfect. Typically in parameter space flavor and CP constraints imply $M_{\rm KK} \gtrsim O(10)$ TeV for the KK threshold \cite{Csaki:2008zd}!

What are we to make of this situation? While flavor and CP tests have very high virtual reach for the warped/composite scenario, they do 
not appear as robust as electroweak constraints. It is indeed plausible that a more refined mechanism for flavor structure is occurring 
within Higgs compositeness so as to relax the bounds significantly, and admit KK states within LHC reach \cite{Fitzpatrick:2007sa}. Because of this, it is imperative that LHC experiments continue to search for KK resonances along the lines of Fig.~\ref{fig:OriginalRS} and \ref{fig:Generalspectrum}, in tandem with ongoing low-energy searches for new sources of flavor and CP violation. But it is also possible that the hierarchy problem is imperfectly solved by Higgs compositeness at a scale $\gtrsim O(10)$ TeV, leaving a {\it Little Hierarchy Problem} between $\sim O(10)$ and $\sim O(1)$TeV. We simply do not understand fundamental physics and the principle of Naturalness underlying the SM hierarchy problem deeply enough to know if they should reliably predict the threshold of new physics to better than a decade in energy. Of course, such a possibility leads to the practical problem that $M_{\rm KK} \gtrsim O(10)$ TeV is 
 outside LHC reach and yet frustratingly close! (It is noteworthy however that such new physics is 
 %
% well 
%
might be
within reach of proposed $100$ TeV colliders).

In this paper, we will pursue the scenario of Higgs compositeness at $\gtrsim O(10)$ TeV. This straightforwardly suppresses all virtual KK-mediated electroweak, flavor and CP violating effects enough to be robustly consistent with all precision experiments to date. 
But we will ask what natural forms of new physics might lie {\it within LHC reach} if we go beyond the minimal structure of Fig.~\ref{fig:OriginalRS} and \ref{fig:Generalspectrum}, without reintroducing conflict with precision tests. We can think of such non-minimal physics lying below the scale at which the hierarchy problem is solved as ``vestiges of naturalness''. If the LHC cannot reach the states central to solving the dominant part of the hierarchy problem (such as KK tops), the search for light vestiges, related to the central players but not among them, are the best hope for the LHC. 

\begin{figure}
\center
%\vspace{-10cm}

%\hspace{-5cm}

\includegraphics[width=0.8\linewidth]{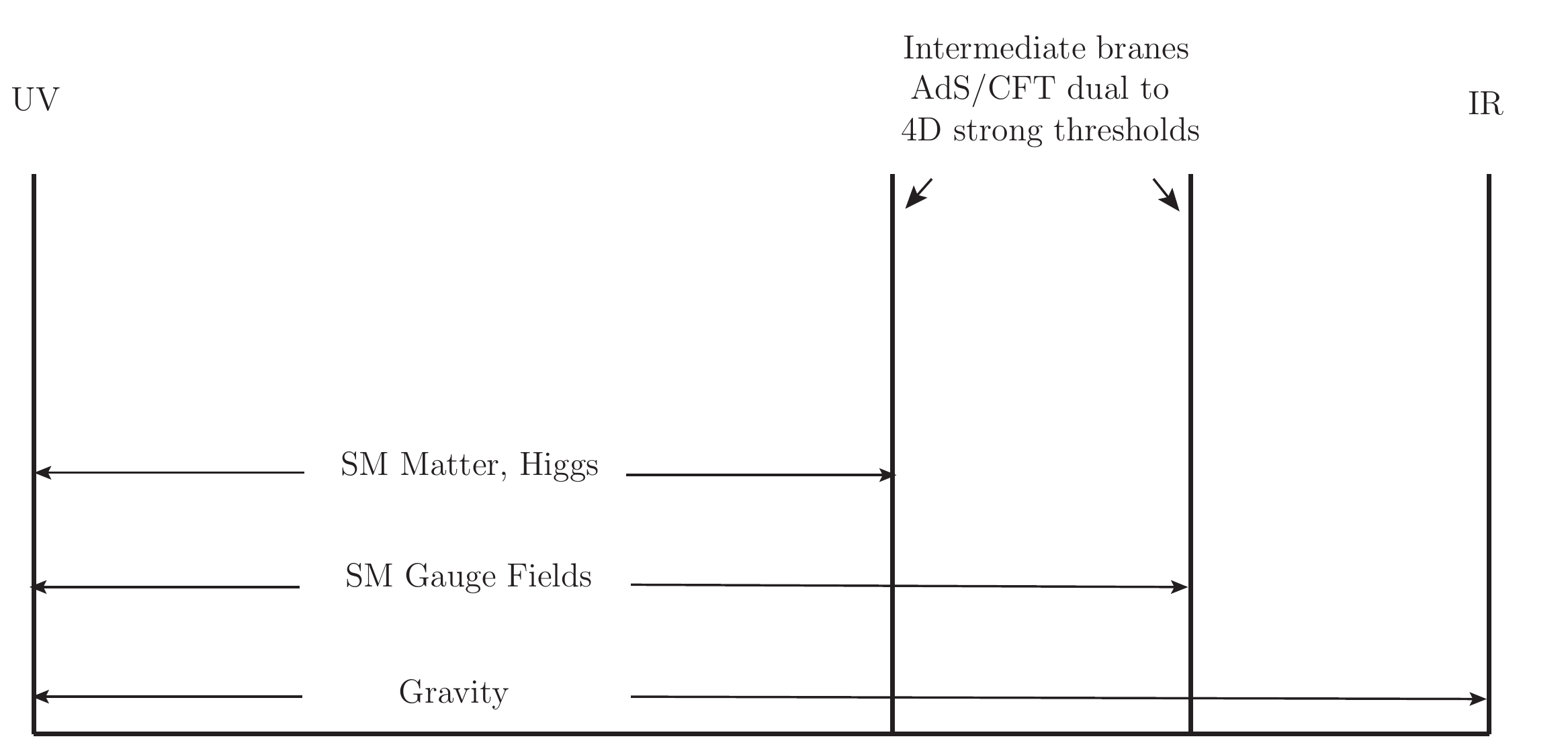}

\caption{Model with two intermediate branes/thresholds.}
\label{fig:Multibrane}
\end{figure}

In particular, we study {\it literally} a straightforward  extension of Fig.~\ref{fig:OriginalRS} which exploits the fact that different types of fields can propagate different amounts into the IR of a warped extra dimension, as schematically depicted in Fig.~\ref{fig:Multibrane}. For simplicity, we focus on three categories of fields: (i) SM matter, including the Higgs, (ii) gauge fields, (iii) gravity. Gravity is the dynamics of all spacetime and therefore must be present in the entire length of the extra dimension in the form of 5D General Relativity. Gauge fields and matter can however reside in a smaller region. Matter fields can live in an even smaller region of the extra dimension than the gauge fields, but not the other way around because charged matter always radiate gauge fields. 
This explains the ordering shown in Fig.~\ref{fig:Multibrane}. The different regions are separated by ``3-branes'', $(3+1)-$dimensional defects in the 5D spacetime. Fig.~\ref{fig:Multibrane} is a simple, robust and interesting generalization of the minimal structure of Fig.~\ref{fig:OriginalRS}, \ref{fig:Generalspectrum}. A quite different proposal using an intermediate brane in warped spacetime was made in \cite{Csaki:2016kqr} in the context of explaining 750 GeV diphoton excess at the LHC \cite{Aaboud:2016tru}.

\begin{figure}
\center

\includegraphics[width=0.6\linewidth]{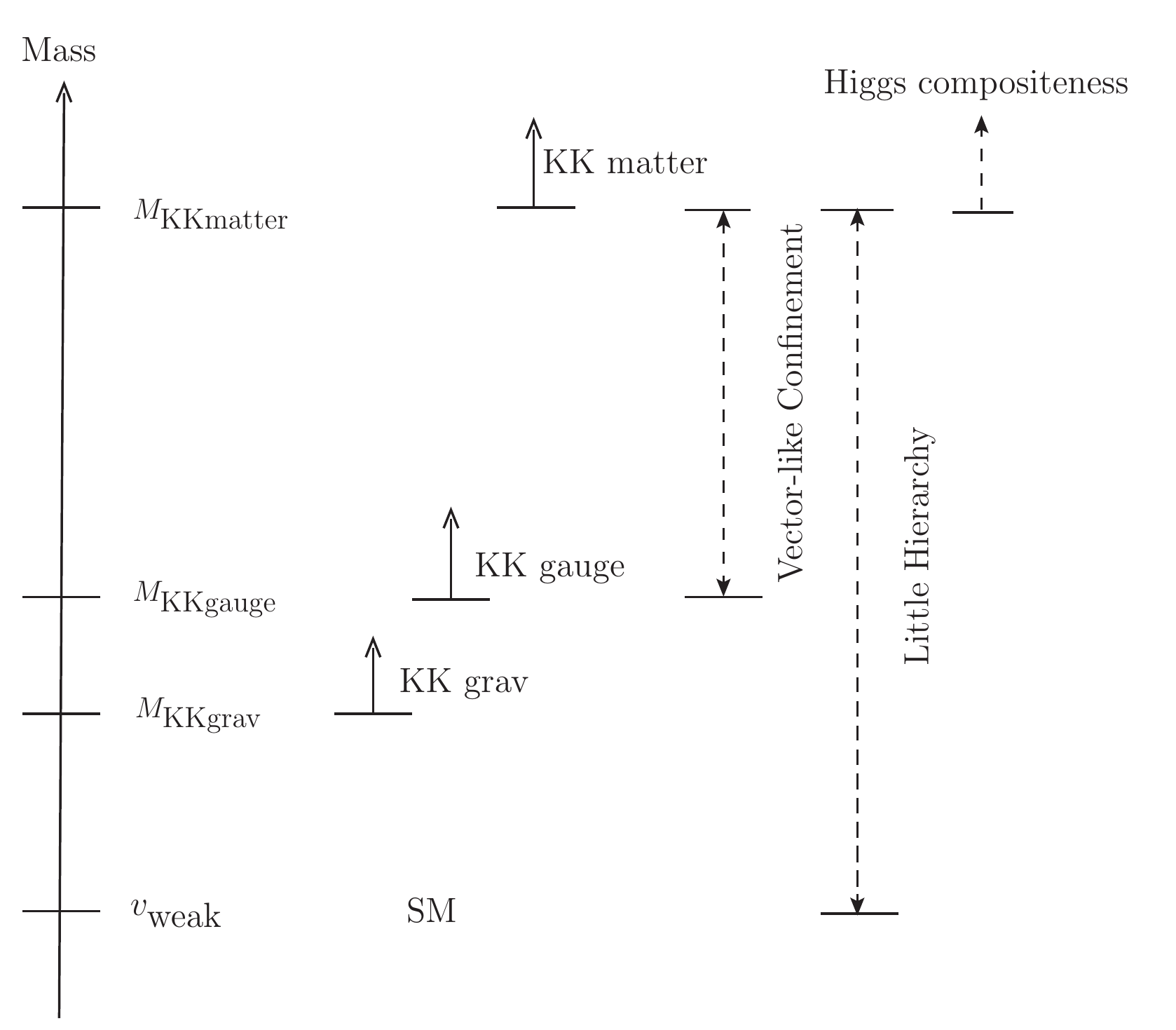}

\caption{Full spectrum of model of Fig.~\ref{fig:Multibrane}.}
\label{fig:fullspectrum}
\end{figure}

The new physics to the IR of Higgs compositeness is (AdS/CFT dual to) that of ``Vectorlike Confinement'', proposed in references \cite{Kilic:2008pm} as a phenomenologically rich structure that is remarkably safe from precision tests, and is a natural candidate for a light vestige of a more general 
dynamics that solves the hierarchy problem. In the framework of Fig.~\ref{fig:Multibrane}, vectorlike confinement incarnates as the extension of the IR of the extra dimension beyond Fig.~\ref{fig:OriginalRS}, resulting in different KK thresholds for matter, gauge fields and gravity as depicted schematically in Fig.~\ref{fig:fullspectrum}. A simple but important result we will demonstrate is that the Goldberger-Wise (GW) mechanism \cite{Goldberger:1999uk} for brane/radion stabilization very naturally results  in ``little'' hierarchies $M_{\rm{KK_{matter,Higgs}}}  \geq M_{\rm{KK_{gauge}}} \geq 
M_{\rm{KK_{grav}}}$. 

From the purely 4D perspective of strong dynamics, the sequence of KK thresholds, $M_{\rm{KK_{matter,Higgs}}}  \geq M_{\rm{KK_{gauge}}} \geq 
M_{\rm{KK_{grav}}}$, is dual to a sequence of strong confinement scales \cite{Raby:1979my, Klebanov:2000hb}, $\Lambda_{\rm{Higgs}} \geq \Lambda_{\rm{meson}} \geq \Lambda_{\rm{glueball}}$. 
Over the large hierarchy from the far UV (the Planck or unification scale) down to $\Lambda_{\rm{Higgs}}$ the strong dynamics is only slowly evolving. At $\Lambda_{\rm{Higgs}}$ the strong dynamics confines ``preons'' into composites, among which is the light SM-like Higgs.
This is analogous to the emergence of pions and heavier hadrons as composites of quarks and gluons upon QCD confinement. But unlike QCD, the strong dynamics does not end at this point, but rather is reorganized into a new set of strongly interacting preons, {\it now approximately decoupled from Higgs and flavor physics}. The IR preons do however carry SM gauge charges. 

At $\Lambda_{\rm{meson}}$ there is a second stage of preon confinement, into ``mesons'' also carrying SM gauge charges. Without direct couplings to the Higgs and SM fermions, this second stage of confinement does not break the SM electroweak chiral symmetries, hence the name ``vectorlike'' confinement. Again, the strong dynamics need not end at this threshold, but can continue with a set of far-IR SM-neutral preons, which ultimately confine into SM-neutral ``glueballs'' at $\Lambda_{\rm{glueball}}$. 

Since the new physics below $\Lambda_{\rm{Higgs}}$ couples to the SM states predominantly via flavor-blind gauge forces, it is naturally safe from the host of electroweak, flavor and CP tests. Phenomenologically, production and decay of the new states below $\Lambda_{\rm{Higgs}}$ will be mediated by on- and off-shell SM gauge bosons. It is very important that experiments search broadly for this kind of physics.
In this way, vectorlike confinement appears as set of ``aftershocks'' of Higgs compositeness, immune to earlier detection but plausibly lying within grasp of the LHC. We will study several aspects of this strongly motivated scenario in this paper. 

\begin{figure}
\center

\includegraphics[width=0.7\linewidth]{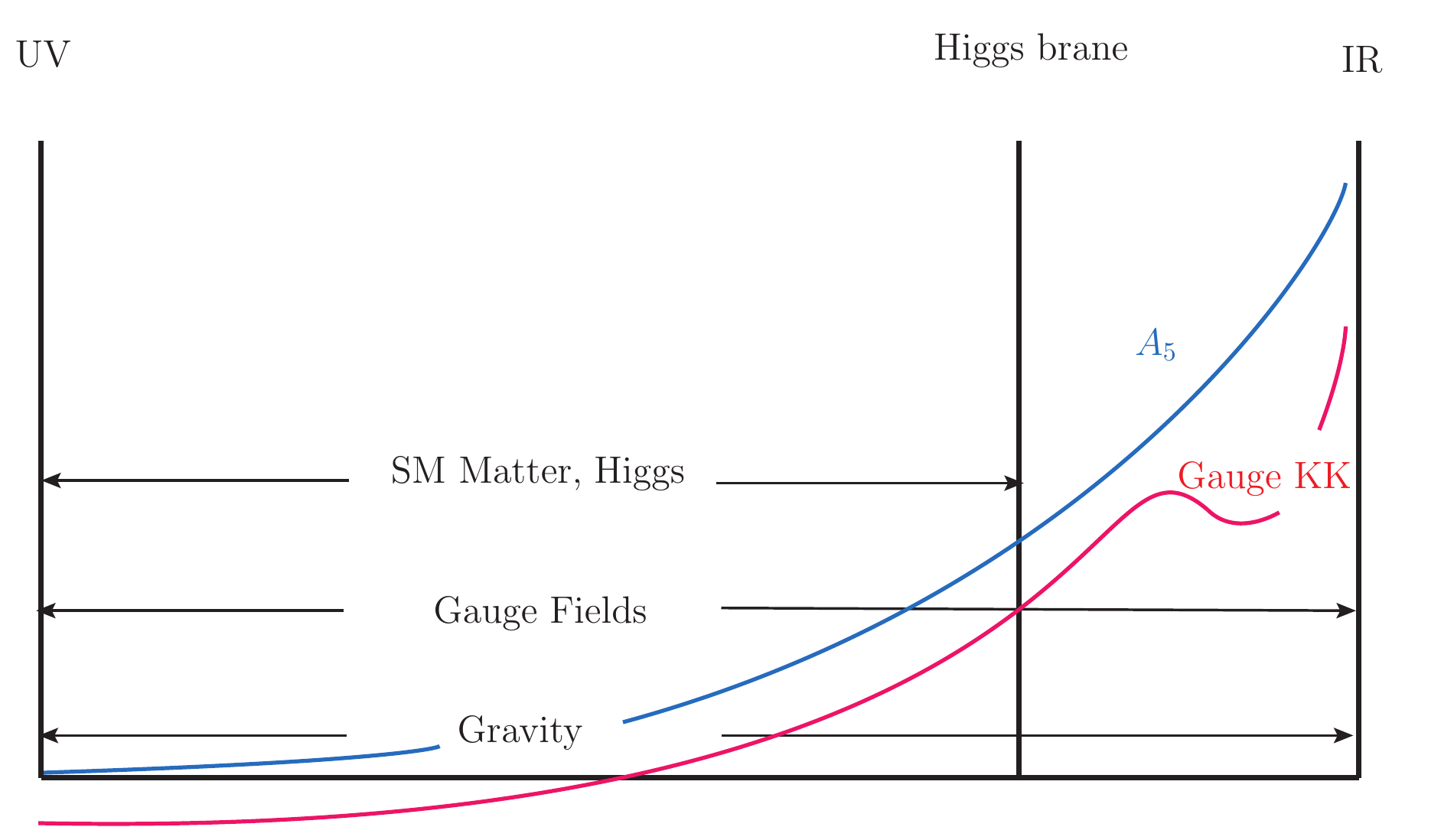}

\caption{Model with an extended gauge group beyond SM and one intermediate brane, resulting in some number of $A_5$ 4D scalars dual to composite PNGB's.}
\label{fig:3brane}
\end{figure}

In references \cite{Kilic:2008pm}, vectorlike confinement was modeled on QCD-like dynamics as the simplest way of illustrating the rich possibilities, using real-world understanding of the strong interactions to stay in non-perturbative theoretical control. A feature of these models is that they typically contain several pseudo Nambu-Goldstone bosons (PNGBs) in the IR of the new physics related to the large chiral symmetry, which can dominate the phenomenology.\footnote{For recent applications of vector-like confinement for explaining the 750 GeV diphoton excess at the LHC, 
%
%\cite{Aaboud:2016tru}, 
%
see, for example, the early references \cite{Franceschini:2015kwy, Nakai:2015ptz}.} However, the specific phenomenological implications are model-dependent. Although QCD-like dynamics do not have a very useful AdS/CFT dual extra-dimensional description, they are in the same ``universality class'' as extra-dimensional models of the type depicted in Fig.~\ref{fig:3brane}, where the 5D gauge group is extended beyond the SM. If UV and IR boundary conditions break some of the gauge symmetry generators, they result in physical extra-dimensional components of the gauge field, ``$A_5$'', which are 4D scalars, AdS/CFT dual to PNGB's \cite{Contino:2003ve}. We will return to study this class of vectorlike confining physics more closely in future work. Unlike in QCD-like constructions, in warped 5D effective field theory we can suppress the existence of $A_5$'s by construction, allowing us to focus on other possibilities for the new phenomenology.

One focus of this paper will be the possibility that lightest new states are the universal ones arising from 5D General Relativity, the scalar ``radion'' measuring the (dynamical) size of the final IR segment of the extra-dimensional interval, and spin-2 KK gravitons. These are the hallmarks of warped extra-dimensional physics. Via the AdS/CFT correspondence these states are dual to special  ``glueballs'' interpolated by the conserved energy-momentum tensor of the strong dynamics, the universal composite operator of any quantum field theory. In particular, this symmetric tensor naturally interpolates spin-2 glueballs dual to KK gravitons, while its Lorentz-trace interpolates the ``dilaton'', a glueball dual to the radion. We will derive and discuss their phenomenological implications, pointing out (i) when they are likely to be the first discovered new states beyond the SM, (ii) their special distinguishing features and the contrast with more QCD-like vectorlike confinement and other beyond-SM physics, (iii) how we can experimentally test whether the new physics is well-described by higher-dimensional dynamics.

In table \ref{table_radion_KK gravity}, we highlight a couple of signals from the gravity sector, namely radion in the model with one intermediate brane of Fig.~\ref{fig:3brane} and KK graviton in the model with two intermediate branes of Fig.~\ref{fig:Multibrane}: further details will be provided in the relevant parts of the paper.
For now, it is noteworthy that the decays in these cases dominantly occur to   
pairs of SM gauge bosons, cf.~top/Higgs playing this role 
in the minimal model of Fig.~\ref{fig:OriginalRS}.
Also, we see that radion and KK graviton are allowed to be lighter than
gauge KK modes\footnote{It might be also possible to make KK graviton lighter than gauge KK using brane-localized kinetic terms (BKT) for gravity \cite{Davoudiasl:2003zt}. For recent applications of this idea for explaining the 750 GeV diphoton excess at the LHC 
%
%\cite{Aaboud:2016tru} 
%
using KK graviton, see,
\cite{Carmona:2016jhr, Falkowski:2016glr}.
However, with too large BKT for gravity, the radion might become a ghost \cite{Davoudiasl:2003zt}.}. %thus having sizable production cross-sections.

A second focus of the paper will be connecting the new physics the LHC can discover to the solution of the hierarchy problem beyond its reach. We will show that low-lying KK modes, though mostly decoupled from the Higgs and flavor, will have subdominant decay channels into $t, h, W_L, Z_L$, the traditional signatures of Higgs compositeness. In this way, the LHC would have a valuable resonance-enhanced ``preview'' of the solution to the hierarchy problem by compositeness, only fully accessible to more energetic future colliders. 
In particular, we find that spin-1 KK {\em gauge} bosons are well-suited for this task.
Note that these are dual to composite vector ``$\rho$'' mesons, which arise as a robust feature in the framework of vector-like confinement also.

A representative sample of the above novel probe of top/Higgs compositeness is shown in table \ref{table_KKZ_KKg}: we will of course explain in later sections how we obtained these
numbers (including assumptions made therein), but let us convey our main message using them for now.
We focus on KK -- excited (dual to composite) $Z$ and gluon, 
%
%as probes of Higgs/top compositeness, 
%
where we fix their mass and coupling to light quarks, hence production cross-section (as shown).
However, decay branching ratios (BR's) to various final states still vary for the {\em same} framework as we vary $\Lambda_{\rm Higgs}$: the left-most column corresponds to the standard composite Higgs model
(i.e., single IR brane/scale, Fig.~\ref{fig:OriginalRS}), whereas right extreme is the flavor-blind limit, i.e., Higgs compositeness scale is decoupled, large $\Lambda_{\rm Higgs}$).
%with middle columns doing the interpolation.
%
Remarkably, we see that decay BR's might be sensitive to $\sim 10-15$ TeV Higgs compositeness scale [in the sense that such values of Higgs compositeness scale can result in $\sim O(1)$ deviations from {\em both} flavor-blind and standard limits], which is the ball park of the generic {\em lower} limit on the Higgs compositeness scale from flavor/CP violation!

\begin{table}[tbp]
\begin{adjustwidth}{-2.2cm}{}
\begin{tabular}{ | c || c || c || c | }
\hline 
%%%%%%%%%%%%%%%%%%%%%%%%%%%%%%%%%%%%%%%
% TABLE I (radion / KK spin-2)
%%%%%%%%%%%%%%%%%%%%%%%%%%%%%%%%%%%%%%%
\multicolumn{4}{|c|}{Radion / KK Graviton} \\
\hline \hline
%\multicolumn{5}{|c|}{$\sigma_{ \rm LHC14} \left( pp \rightarrow \hbox{KK} \; Z \right) \sim 3$ fb for 3 TeV mass and inter-KK coupling of $3$} \\
\multicolumn{2}{|c||}{\backslashbox[50mm]{}{}}
&\makebox[12em]{Radion ($\varphi$) }&\makebox[12em]{KK Graviton}\\ 
\hline\hline

\multicolumn{2}{|c||}{Framework} & one intermediate brane (Fig.~\ref{fig:3brane}) & two intermediate branes (Fig.~\ref{fig:Multibrane}) \\
\hline

\multicolumn{2}{ |c|| }{\multirow{2}{*}{Parameters} } & inter-KK coupling = 3 & inter-KK coupling = 3    \\ \cline{3-4}
\multicolumn{2}{ |c|| }{}  & $M_{\rm KKgauge} = 3$ TeV; $m_{ \varphi 
%
%\rm radion
%
} = 2$ TeV & $M_{\rm KKgauge} =  2$ TeV; $M_{\rm KK grav} = 1$ TeV  \\ \cline{1-4}
\hline \hline

\multicolumn{2}{|c||}{$\sigma_{ \rm LHC14} \left( pp \rightarrow  \hbox{Radion / KK \; Graviton} \right)$} & $\sim 50$ fb & $\sim 100$ fb \\
\hline
%\diaghead(5,-2){Final state}
%{Final state}{$\Lambda_{ \rm Higgs }$}&
%\thead{$3$ TeV \\ (same as spin-1)}&\thead{$10$ TeV}&\thead{ |$ 15$ TeV}&\thead{$\infty$}
%\\ \hline\hline 

\multicolumn{1}{ |c  }{\multirow{4}{*}{BR} } &
\multicolumn{1}{ |c|| }{gg} & $ \sim 90 \% $ & $ \sim 90 \% $    \\ \cline{2-4}
\multicolumn{1}{ |c  }{} & \multicolumn{1}{ |c|| }{ZZ} & $\lesssim 4 \%$ & $\lesssim 4 \%$  \\ \cline{2-4}
\multicolumn{1}{ |c  }{} & \multicolumn{1}{ |c|| }{WW} & $\sim 4 \%$ & $\sim 4 \%$  \\ \cline{2-4}
\multicolumn{1}{ |c  }{} & \multicolumn{1}{ |c|| }{$\gamma \gamma$} & $\sim 0.1 \%$ & $\sim 0.1 \%$  \\ \cline{1-4}
%\multirow{4}{*}{BR} & gg  & $ \sim 90 \% $ & $ \sim 90 \% $ \\
%\tabularnewline
%\hline
%& ZZ & $\lesssim 4 \%$ & $\lesssim 4 \%$ \\
%\tabularnewline
%\hline
%& WW & $\sim 4 \%$ & $\sim 4 \%$ \\
%\tabularnewline
%\hline
%$\gamma \gamma$ & $\sim 0.1 \%$ & $\sim 0.1 \%$ \\
%\tabularnewline

\hline
\end{tabular}
\caption{Estimates for production cross section and decay BR's of radion (left) and KK graviton (right) for a given choice of framework and parameters. For radion, model with one intermediate brane is considered with radion mass 2 TeV, $M_{\rm KK gauge} = 3$ TeV, and inter-KK coupling of 3 [for both 
%
%inter-KK 
%
gluon ($g_{\star}^{\rm QCD}$) and 
%
%inter-KK 
%
gravity ($g_{\star}^{\rm grav}$), which we define in section \ref{subsec: Parameters}]. For KK graviton, we instead considered model with two intermediate branes, in which KK graviton is naturally lighter than KK gauge boson. Similarly to the radion case, inter-KK coupling of 3 [for both $g_{\star}^{\rm QCD}$ and $g_{\star}^{\rm grav}$] is taken.}
\label{table_radion_KK gravity}
\end{adjustwidth}
\end{table}

\begin{table}[tbp]
\begin{adjustwidth}{-0.8cm}{}
\begin{tabular}{ | c || c || c | c || c | }
\hline 
%%%%%%%%%%%%%%%%%%%%%%%%%%%%%%%%%%%%%%%
% TABLE II
%%%%%%%%%%%%%%%%%%%%%%%%%%%%%%%%%%%%%%%
\multicolumn{5}{|c|}{KK Z} \\
\hline \hline
\multicolumn{5}{|c|}{$\sigma_{ \rm LHC14} \left( pp \rightarrow \hbox{KK} \; Z \right) \sim 3$ fb for 3 TeV mass and inter-KK coupling of $3$} \\
\hline

\backslashbox[50mm]{Final state}{$\Lambda_{ \rm Higgs }$}
&\makebox[6em]{$3$ TeV (Fig.~\ref{fig:OriginalRS})}&\makebox[6em]{$10$ TeV}&\makebox[6em]{$ 15$ TeV}
&\makebox[6em]{$\infty$}\\ 
\hline\hline

%\diaghead(5,-2){Final state}%
%{Final state}{$\Lambda_{ \rm Higgs }$}&
%\thead{$3$ TeV \\ (same as spin-1)}&\thead{$10$ TeV}&\thead{$ 15$ TeV}&\thead{$\infty$}
%\\ \hline\hline 
di-leptons $ ( e + \mu ) $ & $ \sim 0 $ & $\gtrsim 6-3 \% $ & $\gtrsim 6 \%$ & $6 \%$
\tabularnewline
\hline
di-bosons (Higgs/$W/Z$) & $79 \%$ & $\sim 0-44 \%$ & $\sim 0 - 7 \%$ & $7 \%$
\tabularnewline
\hline
di-tops & $21 \%$ & $9-15 \% $ & $9-10 \%$ & $10 \%$
\tabularnewline
\hline
di-jets & $\sim 0$ & $63-28 \% $ & $63 -57 \%$ & $57 \%$
\tabularnewline
%\multicolumn{5}{|c|}{Estimates for decay BR's of KK $Z$ for various values of top/Higgs compositeness scale } \\
%\multicolumn{5}{|c|}{($\Lambda_{ \rm Higgs }$), for fixed spin-1 mass scale of 3 TeV and inter-KK $Z$ coupling  ($g^{ Z }_{ \star }$, which we define in}  \\
%\multicolumn{5}{|c|}{section \ref{subsec: Parameters}) of $3$, corresponding to cross-section of $\sim 3$ fb.} \\
\hline\hline\hline
%%%%%%%%%%%%%%%%%%%%%%%%%%%%%%%%%%%%%%%
% TABLE III
%%%%%%%%%%%%%%%%%%%%%%%%%%%%%%%%%%%%%%%
\multicolumn{5}{|c|}{KK Gluon} \\
\hline\hline
\multicolumn{5}{|c|}{$\sigma_{ \rm LHC14} \left( pp \rightarrow  \hbox{KK \; gluon} \right) \sim 200$ fb for 3 TeV mass and inter-KK coupling of 3} \\
\hline
\backslashbox[50mm]{Final state}{$\Lambda_{ \rm Higgs }$}
&\makebox[6em]{$3$ TeV (Fig.~\ref{fig:OriginalRS})}&\makebox[6em]{$10$ TeV}&\makebox[6em]{$ 15$ TeV}
&\makebox[6em]{$\infty$}\\ 
\hline\hline
di-jets (light quarks $+ b$) & $\sim 0 $ & $83-91 \%$ & $86-91 \%$ & $83 \%$
\tabularnewline
\hline
di-tops & $100 \% $ & $ 17-9 \%$ & $14-9 \%$ & $17 \%$
\tabularnewline
\hline
\end{tabular}
\caption{Estimates for decay BR's of KK $Z$ (top) and KK gluon (bottom) for various values of top/Higgs compositeness scale ($\Lambda_{ \rm Higgs }$), for fixed spin-1 mass scale of 3 TeV and inter-KK $Z$/gluon coupling  [$g^{ Z / {\rm gluon} }_{ \star }$, which we define in section \ref{subsec: Parameters}] of $3$, corresponding to cross-section of $\sim 3$ fb (for KK $Z$) and $\sim 200$ fb (for KK gluon).}
\label{table_KKZ_KKg}
\end{adjustwidth}
\end{table}

The paper is organized as follows. 
We begin in section \ref{3brane} with laying out the structure of the model with gauge and gravity propagating in the same bulk, but matter/Higgs
in a subspace, i.e., with the usual UV and IR branes along with a single intermediate brane de-marking the matter/Higgs endpoint.
In section \ref{pheno}, we then describe salient features of the LHC signals of this framework.
In section \ref{4brane}, we discuss more general framework with two intermediate branes, in which gravity extends even beyond the gauge bulk.
% i.e., we have {\em four} branes; this is briefly outlined in section \ref{4brane}, followed by conclusions/outlook.
Section \ref{sec:conclusion} provides our conclusions and outlook.
Some technical details are relegated to the appendices.

\section{Model with {\em one} intermediate brane}

\label{3brane}

\begin{figure}
\centering

\includegraphics[height=65mm]{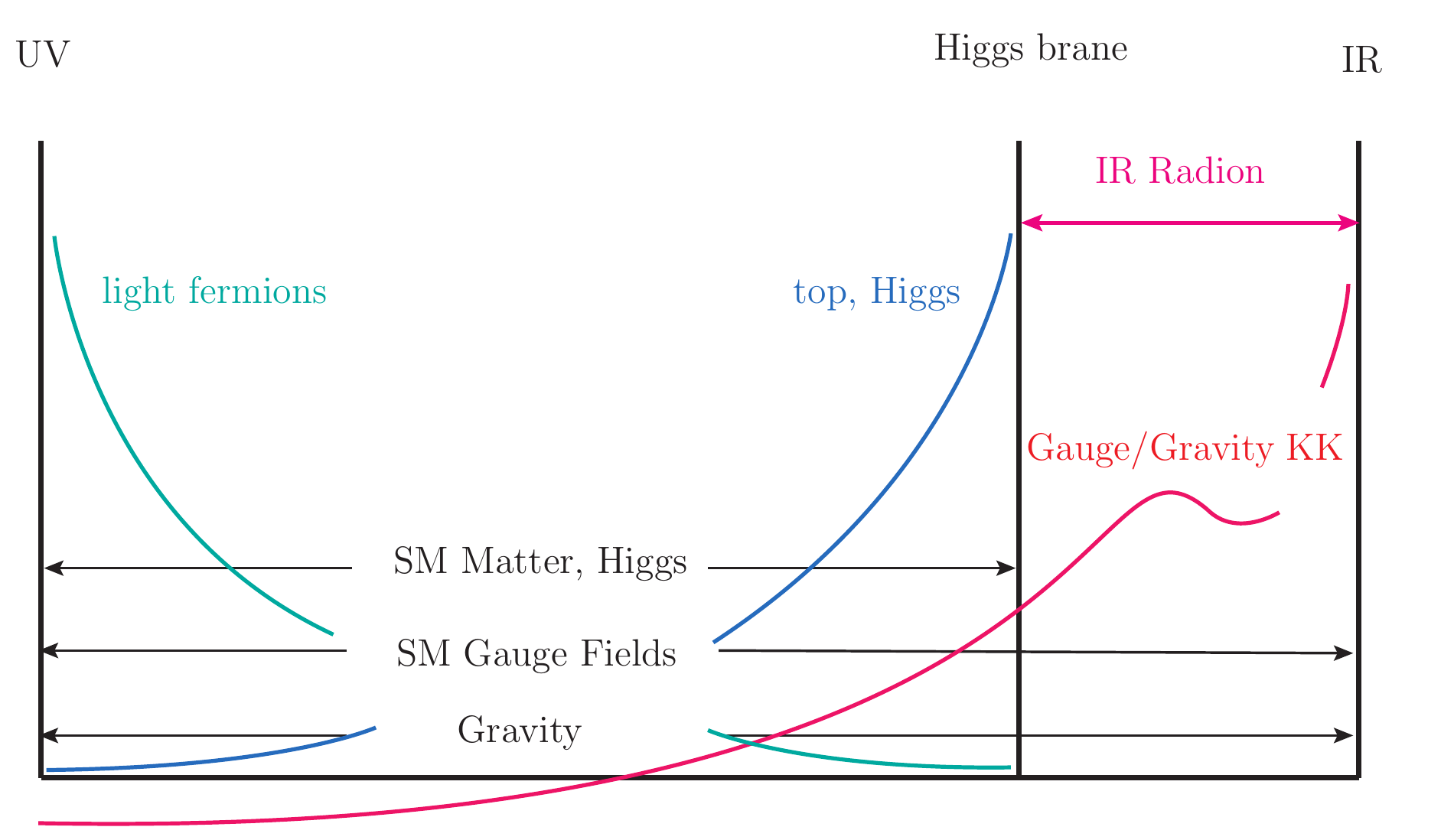}
\caption{Model with one intermediate brane showing light IR radion degree of freedom.}
\label{fig:3brane_w/0_A5}
\end{figure}

We consider gauge and gravity living in the {\em same} bulk starting at the UV brane, with scale
$\Lambda_{ \rm UV } \lesssim M_{ \rm Pl }$ and ending at the IR brane, with scale $\Lambda_{ \rm IR }$, which can be as low as $\sim$ a couple of TeV: see  Fig.~\ref{fig:3brane_w/0_A5}. In the notation used in section \ref{sec: Introduction}, both $\Lambda_{\rm meson}$ 
%
% \sim 
%
and
$\Lambda_{\rm glueball}$ are $\sim  \Lambda_{\rm IR}$, which are also (roughly) the gauge and graviton KK scales 
in the 5D model. 
For now, we will assume the gauge symmetries to be only the SM throughout the bulk so that
we do not have $A_5$'s; we will briefly discuss the latter possibility in section \ref{section:A_5}.
The rest of the SM propagates from the UV brane to an intermediate brane (dubbed ``Higgs'' brane), taken to be $\sim O(10)$ TeV consistently with (anarchic) flavor bounds. We will discuss more details below, showing that even with contribution from composite states of strong dynamics below $\sim \Lambda_{\rm Higgs}$, our framework is indeed safe from EW and flavor/CP violation precision tests. %\footnote{resonances -- whether fermion or gauge KK -- with roughly {\em that} mass, i.e., which are peaked there, will have the usual anarchic/flavor non-universal couplings to SM matter}.
%
%In particular, the top quark(s) and Higgs are peaked at Higgs brane ($\Lambda_{ \rm Higgs }$ scale), implying that (as already alluded to in the introduction) we simply accept meso-tuning of 1 part in $\sim \left( \Lambda_{ \rm Higgs } / 500 \; \hbox{GeV} \right)^2$ where 500 GeV is the mass scale where top quark divergence in 125 GeV SM Higgs mass has to be cut-off for 100\% naturalness (vs.~here this is instead achieved at $\Lambda_{ \rm Higgs }$).
%
As usual, the lighter SM fermions are assumed to be peaked near the UV brane.

%\noindent
%{\bf Language for rest of the paper}

%In the following, we will (mostly) use dual strong dynamics picture (restricting to the lightest composites) of the above 5D model for deducing the spectrum and couplings of this framework (or equivalently the 2-site approach \cite{Contino:2006nn}, i.e., an effective description of the lightest KK modes of the 5D model).

\subsection{Parameters}
\label{subsec: Parameters}

We use the usual notation where $M_5$ is the 5D Planck scale and $k$ is the AdS curvature scale.
The cubic self-coupling of graviton {\em KK} modes (or that of one graviton KK to any two modes localized near IR brane, for example, KK 
gauge) is then 
given (roughly) by
\bea
g^{ \rm grav }_{ \star } & \equiv & \sqrt{ \frac{ k^3 }{ M_5^3 } }
\eea
Also, 
$g^{ \rm grav }_{ \star }$ is dual to coupling of three composites, one of which being spin-2 (and for which we will use the {\em same} notation).

Similarly, $g_5$ is the (dimension{\em ful}) 5D gauge coupling, with the coupling between (three) 4D modes (one of which is gauge KK) localized near IR brane (or three composites, with one being spin-1) given (roughly) by 
\bea
g^{ \rm gauge }_{ \star } & \equiv & g_5 \sqrt{ k } 
\label{g_star_gauge}
\eea
As usual, the sizes of both $g_{ \star }$'s are constrained by perturbativity and fitting observed/4D SM couplings
(i.e., of {\em zero} modes).

However, in the model at hand, there is a {\em new} ingredient, namely, the intermediate (Higgs/matter) brane which
has tension, i.e., is gravitating, resulting in 

\vspace{0.1in}

\noindent
(i) $k$ being different on the two sides of this brane and 

\vspace{0.1in}

\noindent
(ii) a new 
%
%strong coupling scale
%
perturbativity constraint associated with branon (brane-bending) degree of freedom.

We will discuss these issues in detail in Appendix \ref{choice_detail}; here we simply summarize.
The following choices of couplings (in the far IR) suffice for having a 
{\em finite} regime of validity of 5D effective field theory (including the branon degree of freedom):
\bea
g^{ \rm grav}_{ \star \; \rm UV} & < & g^{ \rm grav}_{ \star \; \rm IR }  \lesssim  3 \nonumber \\
g_{ \star \; \rm UV}^{ \rm gauge} & \sim & g_{ \star \; \rm IR }^{ \rm gauge}  \sim  3
\eea
{\em while giving observable signals}.

\subsection{Spectrum}

We expect to have two radions (dual to dilatons in the CFT description), roughly corresponding to fluctuations of Higgs brane relative to UV (heavier mode)
and that of IR brane relative to Higgs brane.
We now work out some of the details of this picture.
We first give a 
%
%quick/
%
schematic review of GW mechanism in CFT language for the minimal model of Fig.~\ref{fig:OriginalRS} \cite{Rattazzi:2000hs}.
We start in the UV with 
\bea
{\cal L} \left( \Lambda_{ \rm UV } \right) & \ni & {\cal L}_{ \rm CFT } + \lambda \; \Lambda_{ \rm UV }^{ \epsilon } \; {\cal O}_{ \rm GW } 
\label{eq:lagrangian_GW_stabilizaiton}
\eea
where ${\cal O}_{ \rm GW }$ is scalar operator 
with {\em scaling} dimension $\left( 4 - \epsilon \right)$ (with $\epsilon > 0$): we also use the convention where 
its naive/engineering dimension is the same so that the coupling constant $\lambda$ above is dimension{\em less}.
We assume that 
${\cal O}_{ \rm GW }$ acquires VEV in the IR, breaking conformal symmetry; this scale can be thought of as VEV of the dilaton field (denoted by $\Phi$ of mass dimension $+1$).
So, we get the dilaton potential  
\bea
{\cal L} \left( \Lambda_{ \rm IR} \right) & \ni & \left( \partial_{ \mu } \Phi \right)^2 + \lambda^{ \prime } \Phi^4 + 
d \lambda \;
\Phi^4 \left( \frac{ \langle \Phi \rangle }{ \Lambda_{ \rm UV } } \right)^{ - \epsilon } 
%+ \mathcal{O} \left( \lambda^2 \left( \frac{ \langle \Phi \rangle }{ \Lambda_{ \rm UV } } \right)^{ - 2 \epsilon } \right)
\label{V_dilaton}
\eea
where we have dropped subleading terms, i.e. $O \left( \lambda^2 \left( \frac{ \langle \Phi \rangle }{ \Lambda_{ \rm UV } } \right)^{ - 2 \epsilon } \right)$, and the second term on the RHS is consistent with conformal symmetry. 
%
%preserving 
%
In the third term, 
%
%we used the dilaton/radion as a ``compensator'' for $\Lambda_{\rm IR}$ 
%
%(or)
%
the IR scale (i.e., RG scaling of perturbation to the CFT) is set by VEV of $\Phi$ (i.e. $\Lambda_{ \rm IR } \sim \langle \Phi \rangle$)
%
%\ka{I think the 2 preceding phrases mean the same?!}
%
[here, we assume that the scaling dimension of ${\cal O}_{ \rm GW }$ remains $( 4 - \epsilon )$ even in the IR],
with $d$ being  an O(1) factor.

Minimizing above potential in the IR, we see that the radius is 
stabilized, i.e., IR scale is fixed as
\bea
\Lambda_{ \rm IR } ( \sim \langle \Phi \rangle ) & \sim & \Lambda_{ \rm UV } \left( -d \frac{\lambda}{\lambda'} \right)^{\frac{1}{ \epsilon }} \\
& \ll & \Lambda_{ \rm UV }, \; {\rm assuming} \; \epsilon < 1
\label{epsilon_value}
\eea
where $\sim$ above (and henceforth) indicates validity up to $O(1)$ factors.
%
%where  $O(1)$ (negative) factor in exponent in 1st line above comes from
%ratio of above $\lambda$'s.
%
In particular (and as is well-known), we see that $\epsilon \sim O(0.1)$ together with $(-d \lambda/\lambda') \sim 
%
%O(0.1)
%
O \Big[ 1 / \left(  \hbox{a few} \right) \Big]
$ 
suffices to generate Planck-weak hierarchy.

Once again, in the model at hand, 
we will have two copies of above module, roughly speaking corresponding to the 
two hierarchies, i.e., $\Lambda_{ \rm IR } / \Lambda_{ \rm Higgs }$ and
$\Lambda_{ \rm Higgs } / \Lambda_{ \rm UV }$.
As shown in more detail in the Appendix \ref{2-dilaton}, the two stabilizations can be done ``sequentially'', giving a heavy dilaton (mass
dictated by $\Lambda_{ \rm Higgs }$) and lighter one (mass $\propto \Lambda_{ \rm IR }$): for the purpose here (i.e., LHC signals), 
we will simply focus on the latter, for which $\Lambda_{ \rm Higgs }$ can be simply
taken to be a ``fixed/UV'' scale.
%
%perhaps with some (mild) tuning, radion corresponding to $\Lambda_{ \rm IR }$
%can be made (much) lighter
%than its going-rate (vs.~keep heavier radion at its more natural value) so that 
%2 radions decouple, even more than naively expected, i.e., based simply on ratio of $\Lambda$'s (?)...
%
The physical dilaton (denoted by $\varphi$) corresponds to fluctuations around VEV, i.e.,
\bea
\Phi \sim \Lambda_{ \rm IR }  + a g^{ \rm grav }_{ \star } \varphi
\label{dressing}
\eea
where $a$ is an $O(1)$ factor.
Plugging this into the above potential, the {\em lighter} dilaton  
mass is then given by \cite{Konstandin:2010cd, Chacko:2012sy}
\bea
m^2_{ \varphi } & \sim & \epsilon \; \lambda' \; \Lambda_{ \rm IR }^2 
%
%m_{ \rm gauge }^2
%
\label{radion_mass}
\eea
where $\epsilon$ is then (roughly) set (as above) to logarithm of hierarchy (the one relevant here is between Higgs and IR branes) and
$\lambda'$ is dual, in 5D, to the amount of detuning of IR brane tension.
%
%i.e., the -- relative -- deviation of the IR brane tension from its tuned value.
%
So, to summarize the various scales, we consider the case:
%
%\bea
%m_{ \varphi }  & \lesssim m_{ \rm gauge \; KK/composite} \sim m_{ \rm grav \; KK/composite } & %\sim \Lambda_{ \rm IR }
%\eea
%
\bea
m_{ \varphi } \lesssim 
%
%\Lambda_{\rm meson} \sim \Lambda_{\rm glueball} \sim 
%
\Lambda_{\rm IR} \ll \Lambda_{ \rm Higgs }.
\eea

\subsection{Couplings}

\subsubsection{Radion/Dilaton}

Once again, we treat the separation between UV and Higgs brane to be fixed, thus reducing the
(light) radion/dilaton analysis to the usual minimal case with only two branes.
We then simply drop the label ``IR'' on dilaton and ${\cal O}_{ \rm GW }$. 

\noindent 
{\bf Coupling to SM gauge fields}

These can be deduced from the running of the SM gauge couplings as follows.
We start with value $g_{ \rm UV }$ at $\Lambda_{ \rm UV }$ and pass through various thresholds all the way to $M_Z$ \cite{Agashe:2005vg}:
\bea
\frac{1}{ g_{ \rm SM }^2 } & \approx & \frac{1}{ g_{ \rm UV }^2 } + 
%
%\frac{N}{ 16 \pi^2 } 
%
%\frac{1}{ \left( g^{ \rm gauge }_{ \star } \right)^2 }
%
b_{ \rm strong \; UV } 
\log \left( \frac{ \Lambda_{ \rm UV } }{ \Lambda_{ \rm Higgs } } \right) 
+ b_{ \rm strong \; IR } 
\log \left( \frac{ \Lambda_{ \rm Higgs } }{ \Lambda_{ \rm IR } } \right) 
+ \nonumber \\ 
%
%\frac{ 
%
& & \left( b_{ \rm SM } - b_{ \rm top, \; Higgs } \right)
%
%}{ 16 \pi^2 } 
%
\log \left( \frac{ \Lambda_{ \rm UV } }{ \Lambda_{ \rm Higgs } } \right)
+
b_{ \rm SM } 
\log \left( \frac{ \Lambda_{ \rm Higgs } }{ M_Z } \right)
\label{gauge_match}
\eea
where $b_{ \rm strong \; UV \; (IR) }$ 
%
%1 / \left( g^{ \rm gauge }_{ \star } \right)^2 \sim N_{ \rm CFT } / \left( 
%16 \pi^2 \right)$ 
%
are the contributions of UV and IR 4D strong dynamics (including, in the former case, 
the SM top quark and Higgs, which are composites), respectively, to the running of the SM gauge coupling
and $b_{ \rm SM }$ is the usual SM contribution.

We expect 
\bea
b_{ \rm strong } & = & \frac{ O \left( N_{ \rm strong } \right) }{ 16 \pi^2 } \nonumber \\
& \sim & \frac{1}{ { \left( g_{ \star }^{ \rm gauge } \right)^2 } }
\label{bCFT}
\eea
where in second line, 
we have used the standard large-$N$ relation that coupling of three composites, i.e., $g^{ \rm gauge }_{ \star }$ (in this case, one being
spin-1/gauge) is given by $\sim 4 \pi / \sqrt{ N_{ \rm strong }}$.
In fact, the 5D result is:
\bea
b_{ \rm strong } & = & \frac{1}{ g_5^2 k } 
\eea
which (as expected) is a good 
match to the second line of Eq.~(\ref{bCFT}) above [using Eq.~(\ref{g_star_gauge})].

The dilaton can be considered to be fluctuations around TeV scale, i.e., $\Lambda_{ \rm IR } \rightarrow \Lambda_{ \rm IR }
+ a g_{ \star }^{ \rm grav } \varphi$ [see Eq.~(\ref{dressing})].
We plug this into the gauge field kinetic term in the form $F_{ \mu \nu } F^{ \mu \nu }/ \left( 4 \; g_{ \rm SM }^2 \right)$,
with $g_{ \rm SM }$ as in Eq.~(\ref{gauge_match}). 
We thus get  
the dilaton coupling to SM gauge bosons \cite{Chacko:2012sy, Csaki:2007ns,Chacko:2014pqa}:
\bea
\delta {\cal L} & \sim & 
g_{ \rm SM }^2 b_{ \rm strong } 
\varphi F_{ \mu \nu } F^{ \mu \nu } \frac{ g^{ \rm grav }_{ \star } }{ \Lambda_{ \rm IR } } +...
\nonumber \\
& \sim & \left( \frac{  g_{ \rm SM } } { g^{ \rm gauge }_{ \star \; \rm IR } } \right)^2 \varphi F_{ \mu \nu } F^{ \mu \nu } \frac{ g^{ \rm grav }_{ \star } }{ \Lambda_{ \rm IR }.
%
%m_{ \rm gauge } 
%
} 
\label{coupling_dilaton_gauge}
\eea
%
%where $g^{ \rm grav }_{ \star }$ is the composite graviton coupling
%
%where $a$ is an $O(1)$ parameter/factor...
%
%
%It is noteworthy that in the unified case, i.e.,
%SM gauges a subgroup of {\em simple} global symmetry group of CFT
%(so that $g^{ \rm gauge }_{ \star }$ is 
%{\em universal}), we get
%
%\bea
%\frac{ \hbox{dilaton coupling to 2 SM gauge bosons} } { \left(\hbox{respective SM gauge coupling}\right)^2 } & \approx & \rm universal , 
%\label{univ_dilaton}
%\eea
%

\noindent
{\bf Coupling to top quark/Higgs}

For simplicity, we assume that the top quark/Higgs are {\em strictly} localized on the
Higgs brane, which (as already mentioned) we are treating (effectively) as ``UV'' brane for
the purpose of obtaining couplings of the light radion.
In the 5D model, we can couple the Higgs and top quarks to the 5D GW field (used for stabilization) evaluated at the Higgs brane, thereby generating a coupling of radion to the top quark/Higgs.
%
%Goldberger, Wise
%
We will work out the size of this induced coupling in the compositeness picture, the
above coupling in the 5D model being dual to:  
\bea
\delta {\cal L} \left( \Lambda_{ \rm Higgs } \right) \sim \frac{ \kappa \Lambda_{ \rm Higgs }^{ \epsilon } }{ \Lambda_{ \rm Higgs }^4 } 
{\cal O}_{ \rm GW } {\cal O}_{ t/ H } 
\eea
where ${\cal O}_{ t/ H }$ is an operator (of mass dimension 4) containing top quark and Higgs fields
(to be discussed more below).
Since ${\cal O}_{ \rm GW }$ obtains a VEV at scale $\Lambda_{ \rm IR }$ (fluctuations around which
correspond to the dilaton), we can interpolate it in the IR as
\bea
{\cal O}_{ \rm GW } & \sim & \Lambda_{ \rm IR }^{ 3 - \epsilon } g^{ \rm grav }_{ \star \; \rm IR } \varphi
\eea
i.e., 
(as above) we can choose derivatives to {\em not} appear on $\varphi$, which implies that
we must allow the most general form of ${\cal O}_{ t/ H }$ (i.e. we can{\em not} integrate by parts
to get rid of derivatives on top quark and Higgs fields):
\bea
{\cal O}_{ t/ H } 
& \ni \bar{t} \; \partial \! \! \! \! \! \!  \not \: \: \; t - \left( \partial^{ \mu } \bar{t} \right) \gamma_{ \mu } t + 
c_1 y_{ t } \bar{t} t H +  c_2 \left( \partial_{ \mu } H^{ \dagger } \right) \partial^{ \mu } H  + c_3 H^{ \dagger } \Box H +
\frac{ \left( y_t, g_{ \rm EW } \right)^2 \Lambda_{ \rm Higgs }^2 }{ 16 \pi^2 }
H^{\dagger } H \nonumber \\
\eea
where $c$'s are {\em independent}/{\em arbitrary} coefficients. %(cf.~in $T^{ \mu \; (t/H)}_{ \mu }$ below).

Let us consider dilaton decay from each term in turn.
A quick, {\em explicit} computation shows that amplitude for $\varphi \rightarrow \bar{t} t$ from the top quark ``kinetic''\footnote{quotes are used
here since these are actually 
multiplied by $\varphi$.} term 
in ${\cal O}_{ t/ H }$ is $\propto m_t$: 
a simple argument based on angular momentum conservation for scalar decay into a fermion-antifermion pair shows that it must be so.
So, the first two terms actually contribute similarly to the third term, i.e., ``mass'' term
(where we have included $y_t$, i.e., SM top Yukawa, as flavor spurion in the power counting).

On the other hand, for $\varphi \rightarrow H^{ \dagger } H$, i.e., decay into scalars, there is no such constraint from angular
momentum conservation: indeed, we explicitly find that kinetic term for $H$ gives amplitude $\propto p_{ H \; 1 } . p_{ H \; 2 }
\approx m_{ \varphi }^2 /2$ (in the limit of $m_H \ll m_{ \varphi }$).
Note that contribution of the 
$\Box H$ term (for on-shell $H$) {\em is} $\propto m_H^2$, i.e., {\em actual} mass term, which is $\ll m_{ \varphi }^2 $, thus is 
sub-dominant to the Higgs kinetic term.
In the last term in $\mathcal{O}_{ t/ H }$, we have assumed that the SM Higgs complex doublet $H$ is a PNGB so that
its ``mass squared'' is  SM loop factors smaller than $\Lambda_{ \rm Higgs }^2$, i.e., given our choices of
$\Lambda_{ \rm Higgs } \sim O(10)$ TeV and $\Lambda_{ \rm IR } \sim$ a few TeV, we see that this contribution is 
-- roughly and numerically -- comparable to that from the 
Higgs kinetic term.

So, we can just keep top quark {\em mass} and Higgs {\em kinetic} terms in ${\cal O}_{ t/ H }$ above.
We then get 
\bea
\delta {\cal L} \left( \Lambda_{ \rm IR } \right) \sim 
\kappa \left( \Lambda_{ \rm Higgs } \right) \left( \frac{ \Lambda_{ \rm IR } }{ \Lambda_{ \rm Higgs } } \right)^{ 4 - \epsilon } g^{ \rm grav}_{ \star \; \rm IR }
\frac{ \varphi }{ \Lambda_{ \rm IR } } \Big[  m_t \bar{t} t + \left( \partial_{ \mu } H \right)^{\dagger} \partial^{ \mu } H  \Big]
\label{radion_topHiggs}
\eea
which gives a (much) smaller decay width for dilaton into top/Higgs as compared to SM gauge bosons in final state.\footnote{We have checked that other possible
contributions to the radion couplings to top/Higgs are comparable to or smaller than the above.} 

We conclude 
from the above analyses that the production of the radion/dilaton is dominated by gluon fusion; dilaton 
decays mostly to two SM gauge bosons, all via Eq.~(\ref{coupling_dilaton_gauge}).

\subsubsection{Spin-1/Gauge KK}

We focus here on the lightest spin-1 composite, denoted by $\tilde{ \rho }
%
%\hbox{ \small{comp}
%
%^{ \rm comp }
%
$ (reserving $\rho$ for the mass
eigenstate: see below).

\noindent {\bf Flavor universal coupling} 

\begin{figure}
\centering

\includegraphics[height=60mm]{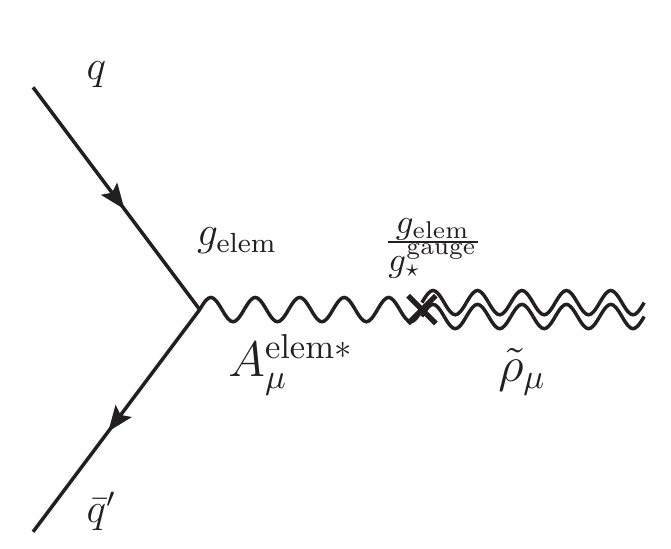}

\caption{Universal spin-1 couplings via elementary-composite mixing (generalization of well-known $\gamma-\rho$ mixing).}
\label{photon_rho}
\end{figure}

The flavor universal part of coupling of $\rho$ (to matter/Higgs fields) is given by a generalization of the well-known phenomenon of 
$\gamma-\rho$ mixing from QCD \cite{Contino:2006nn} (see also Fig.~\ref{photon_rho}), which we briefly
review here.

We start with the kinetic and mass terms  
\bea
{\cal L} & \ni & 
\frac{1}{4} \left( \tilde{ \rho }^{  \mu \nu } \tilde{ \rho }_{ \mu \nu }  +  F^{ \rm elem \; \mu \nu } F_{  \rm elem \; \mu \nu }  \right) + \nonumber \\
& & \Lambda_{ \rm IR }^2 \Big[ \tilde{ \rho }_{ \mu } \tilde{ \rho }^{ \mu } + \frac{ g_{ \rm elem } }{ g_{ \star }^{ \rm gauge } } 
\tilde{ \rho }^{ \mu } A^{ \rm elem }_{ \mu }
+ \left( \frac{ g_{ \rm elem } }{ g_{ \star }^{ \rm gauge } } \right)^2 A^{ \rm elem }_{ \mu } A^{ \rm elem \; \mu } \Big] + \nonumber \\
& & g_{ \rm elem } \bar{q} A^{ \rm elem }_{ \mu } \gamma^{ \mu } q + g_{\star}^{\rm gauge}  \bar{\psi} \tilde{\rho}_{ \mu } \gamma^{ \mu } \psi
\label{lagrangian_rho_photon_mixing}
\eea
where $A^{ \rm elem }_{ \mu }$ denotes gauge field {\em external} to the 4D strong dynamics (thus often called ``elementary''):
{\em all} SM matter (fermions and Higgs boson, denoted generically by $q$ above) couple to it with strength $g_{ \rm elem }$. Similarly, all composite fermions of strongly coupled sector are denoted by $\psi$ and composite vector meson $\tilde{\rho}_{\mu}$ couples to them with strength $g_{\star}^{\rm gauge}$. Note that the second term in the second line of Eq.~(\ref{lagrangian_rho_photon_mixing}), $\sim \tilde{\rho}_{\mu} A_{\mu}^{\rm elem}$, is obtained by starting from $A_{\mu}^{\rm elem} J_{\rm strong \; IR}^{\mu}$ and then using the usual interpolation for (the lightest) spin-1 composites ($\rho$ mesons):
\bea
J_{\rm strong \; IR}^{\mu} \sim \frac{\Lambda_{\rm IR}^2}{g_{\star}^{\rm gauge}} \rho^{\mu}_{\rm IR}.
\label{eq:interpolation_J_rho}
\eea
As we will see, even though the above mass terms break {\em elementary} gauge symmetry, there {\em is} a 
residual gauge invariance (corresponding to a massless field) which we identify with the final SM gauge 
symmetry \cite{Contino:2006nn}. %\sh{I added the following sentence, but we are good to remove it if that's better :)} In fact, it is the masslessness of the SM gauge boson (in mass basis) that fixes the coefficient of the third term in the second line of Eq.~(\ref{lagrangian_rho_photon_mixing}).
We
diagonalize the mass terms by defining the physical states (admixtures of $\tilde{ \rho }$
and $A^{ \rm elem}_{ \mu }$):
\bea
A_{ \mu } & = & \cos \theta A^{ \rm elem }_{ \mu } + \sin \theta \tilde{ \rho }_{ \mu } \\
\rho_{ \mu } & = &  \cos \theta  \tilde{ \rho }_{ \mu } - \sin \theta A^{ \rm elem }_{ \mu }
\label{eigen}
\eea
%
%where $A_{ \mu }$ is the putative (massless) SM gauge field 
%
and 
\bea
\sin \theta &  = & \frac{ g_{ \rm elem } }{ \sqrt{ g_{ \rm elem }^2 + g^{ \rm gauge \; 2 }_{ \star } } }
\eea
This gives
\bea
{\cal L } & \ni & \frac{1}{4} \left( \rho^{ \mu \nu }  \rho_{ \mu \nu }  +  F^{ \mu \nu } F_{ \mu \nu }  \right) 
+  \nonumber \\ 
& & \Lambda_{ \rm IR }^2 \rho^{ \mu } \rho_{ \mu } + g_{\rm SM} \bar{\psi} A^{\mu} \gamma_{\mu} \psi + g_{\star}^{\rm gauge} \bar{\psi} \rho^{\mu} \gamma_{\mu} \psi + \nonumber \\
& & g_{ \rm SM } \bar{q} A^{ \mu } \gamma_{ \mu } q + \frac{ g_{ \rm  SM } g_{ \rm elem } }{ g_{ \star }^{ \rm  gauge } } \bar{q} \rho^{ \mu } \gamma_{ \mu } q + \cdots
\label{coupling_rho}
\eea
where the last term is the (universal) coupling of SM fermions to $\rho$.
Also, as anticipated above, $A_{ \mu }$ is massless (thus corresponds to the SM gauge field), with
\bea
g_{ \rm SM } & = & \frac{ g_{ \star }^{ \rm gauge } g_{ \rm elem } }{ \sqrt{ g_{ \rm elem }^2 + g^{ \rm gauge \; 2 }_{ \star } } }
\eea
being the SM gauge coupling.
Henceforth, we will assume $g_{ \rm elem } \ll g_{ \star }^{ \rm gauge }$ so that 
\bea
g_{ \rm SM } & \approx & g_{ \rm elem }
\eea
and coupling of SM fermions to $\rho$ is $\approx g_{ \rm  SM }^2 /  g_{ \star }^{ \rm  gauge }$.

%
%-- 5D gives
%
%\bea
%b & = & ...
%\eea
%

\noindent {\bf Couplings to radion/dilaton} 

As discussed above, couplings of dilaton/radion can be obtained by using it as a ``compensator'' for $\Lambda_{ \rm IR }$, giving
Eq.~(\ref{coupling_dilaton_gauge}) from dependence of $g_{ \rm SM }$ \footnote{More precisely, dependence of $g_{\rm SM}$ on $\Lambda_{\rm IR}$ originates from dependence of $g_{\rm elem}$ on $\Lambda_{\rm IR}$ via the relation $\frac{1}{g_{\rm SM}^2} = \frac{1}{g_{\rm elem}^2} + \frac{1}{g_{\star}^{{\rm gauge} \; 2}}$.} on $\Lambda_{ \rm IR }$ (via RG evolution of the
gauge coupling) {\em and} 
a coupling to two $\tilde{ \rho }$'s (which gets converted mostly into two $\rho$'s):
\bea
\delta
{\cal L} & \sim & \Lambda_{ \rm IR }^2 \tilde{ \rho }_{ \mu } \tilde{ \rho }^{ \mu } \nonumber \\
& \rightarrow  & \Phi_{ \rm IR }^2  \tilde{ \rho }_{ \mu } \tilde{ \rho }^{ \mu } \nonumber \\
& \ni & 
g^{ \rm grav }_{ \star \; \rm IR } \Lambda_{ \rm IR } \varphi 
\rho^{ \mu } \rho_{ \mu }
\label{dilaton_rho}
\eea
which however is not relevant for collider signals. 
%
%It may be worth mentioning that the naive couplings of $\varphi$ to $A_{\mu}A^{\mu}$ or $A_{\mu} \rho^{\mu}$ one may find above %using $\gamma-\rho$ mixing, after properly adding contributions from the other two terms in the second line of 
%Eq. (\ref{lagrangian_rho_photon_mixing}), vanish.
%
Note that using $\gamma-\rho$ mixing in first line of Eq.~(\ref{dilaton_rho}), one {\em naively} obtains couplings of $\varphi$ to $A_{\mu}A^{\mu}$ or $A_{\mu} \rho^{\mu}$; however, after properly adding contributions from the other two terms in the second line of 
Eq. (\ref{lagrangian_rho_photon_mixing}), we can see that these terms vanish.

In addition, {\em after} radius stabilization/explicit breaking of conformal symmetry, we get a mixed coupling of dilaton, i.e., 
to $\rho$ and SM gauge field as follows.
In the IR, we can interpolate the Goldberger, Wise operator as 
\bea
{\cal O}_{ \rm GW } & \ni & \Lambda_{ \rm IR }^{ -\epsilon } \tilde{ \rho }^{ \mu \nu } \tilde{ \rho }_{ \mu \nu }.
\eea
Plugging above in Eq.~(\ref{eq:lagrangian_GW_stabilizaiton}), RG-running down to $\Lambda_{\rm IR}$ and then promoting $\Lambda_{ \rm IR } \rightarrow \Lambda_{ \rm IR } + a g^{ \rm grav }_{ \star  \; \rm IR} \varphi$, we get
\bea
\delta {\cal L} \left( \Lambda_{ \rm IR } \right) & \sim & \lambda 
%
%\left( \Lambda_{ \rm UV } \right) 
%
\left( \frac{ \Lambda_{ \rm IR } }{ \Lambda_{ \rm Higgs } } \right)^{ - \epsilon }  \tilde{ \rho }^{ \mu \nu } \tilde{ \rho }_{ \mu \nu }
\nonumber \\
& \sim & \lambda 
%
%\left( \Lambda_{ \rm UV } \right) 
%
\epsilon g^{ \rm grav }_{ \star \; \rm IR } \left( \frac{ \Lambda_{ \rm IR } }{ \Lambda_{ \rm Higgs } } \right)^{ - \epsilon } \frac{ \varphi }{ \Lambda_{ \rm IR } }  
\tilde{ \rho }^{ \mu \nu } \tilde{ \rho }_{ \mu \nu }.
\eea
Finally, plugging the mass eigenstates from Eq.~(\ref{eigen}) into above gives\footnote{The same procedure also results in couplings of the form $\varphi \rho^{\mu\nu} \rho_{\mu\nu}$ or $\varphi F^{\mu\nu} F_{\mu\nu}$, i.e., 
{\em corrections} to the couplings of dilaton/radion to {\em pairs} of SM gauge fields {\em from radius stabilization} 
\cite{Chacko:2012sy, Chacko:2014pqa} and $\rho$'s;
however, these are {\em sub}-dominant to the pre-existing ones, hence we will not discuss them further.
%
%However, we will not discuss further on these couplings.
%
}:
\bea
\delta {\cal L } \left( \Lambda_{ \rm IR } \right) & \sim & \lambda \left( \Lambda_{ \rm Higgs } \right)  \epsilon g^{ \rm grav }_{ \star \; \rm IR} \left( \frac{ \Lambda_{ \rm IR } }{ \Lambda_{ \rm Higgs } } \right)^{ - \epsilon }
\frac{ g_{ \rm elem } } { g^{ \rm gauge }_{ \star \; \rm IR } } \rho^{ \mu \nu } F^{ \mu \nu } 
\frac{ \varphi}{ \Lambda_{ \rm IR } }.
\label{dilaton_rho_photon}
\eea
From Eq.~(\ref{epsilon_value}), here we have $\epsilon \sim 1 / \log \left( \Lambda_{ \rm Higgs } / \Lambda_{ \rm IR } \right) \sim 1 / $ a few, since
the relevant hierarchy is $\Lambda_{ \rm Higgs } / \Lambda_{ \rm IR }$ as indicated, not $\Lambda_{ \rm UV } / \Lambda_{ \rm IR }$, and from this we also see that $\left( \Lambda_{ \rm IR } / \Lambda_{ \rm Higgs } \right)^{ - \epsilon }$ is an $O(1)$ factor. 
Thus, the $\rho$-dilaton-SM gauge boson coupling in Eq.~(\ref{dilaton_rho_photon}) can be (roughly) comparable to the last term in
Eq.~(\ref{coupling_rho}), i.e., universal $\rho$ coupling (assuming $g^{ \rm grav }_{ \star \; \rm IR } \sim 1$).
Note that 
{\em decay} of 
%
%there is {\em no} coupling of 
%
$\rho$ to two $\varphi$ (cf.~spin-2 below) is {\em not} allowed by a combination of Bose-Einstein statistics and 
angular momentum conservation arguments. 
%
%\ka{I went with the ``safe'' route here...}
%
%\ka{actually, is there a principle behind this...
%or does it suffice to argue that {\em decay} of {\em on}-shell 
%$\rho$ to 2 (identical) scalars is not {\em allowed} by Bose-Einstein statistics combined
%with angular momentum conservation?!}
%

\noindent {\bf Flavor {\em non}-universal couplings to top/Higgs} 

On the other hand, the flavor non-universal part of the $\rho$ couplings 
(relevant only for top quark/Higgs: negligible for light fermions, at least for LHC signals) arises from 
\bea
\delta {\cal L} \left( \Lambda_{ \rm Higgs } \right) \sim
\frac{ \left( g^{ \rm gauge }_{ \star \; \rm UV } \right)^2 }{ \Lambda_{ \rm Higgs }^2 } 
J^{ \mu }_{ \rm strong \; IR } \left( \bar{t} \gamma_{ \mu } t + H^{ \dagger } D_{ \mu } H \right) 
\label{eq:lagrangian_J_CFT IR_to_t/H}
\eea
where %we have assumed that 
this coupling of top/Higgs to IR strong dynamics is generated by integrating out physics of top/Higgs compositeness at scale
%exchange of spin-1 fields of mass
$\sim \Lambda_{ \rm Higgs }$, with a coupling characteristic of gauge sector of the UV strong dynamics (see appendix \ref{2-dilaton} for further explanation of the UV and IR CFT's with stabilization mechanism).
This runs down to the IR:
\bea
\delta {\cal L} \left( \Lambda_{ \rm IR } \right) & \sim & \frac{ \left( g^{ \rm gauge }_{ \star \; \rm UV } \right)^2 }{ g^{ \rm gauge}_{ \star \; \rm IR} } 
\left( \frac{ \Lambda_{ \rm IR } }{ \Lambda_{ \rm Higgs } } \right)^2 \rho^{ \mu }_{ \rm IR }  \left( \bar{t} \gamma_{ \mu } t + H^{ \dagger } D_{ \mu } H \right) 
\label{gauge_topHiggs}
\eea
\noindent where we have used the interpolation relation of Eq.~(\ref{eq:interpolation_J_rho}). 

Clearly, the production of $\rho$ at the LHC proceeds via light quark coupling in last term in Eq.~(\ref{coupling_rho}), while 
decays occur via same coupling {\em and} that in Eq.~(\ref{gauge_topHiggs}) and~(\ref{dilaton_rho_photon}), assuming
$\varphi$ is lighter than $\rho$.

%using the usual interpolation for spin-1 composites  ($\rho$ mesons):
%
%\bea
%J^{ \mu }_{ \rm CFT \; IR } & \sim & \frac{ \Lambda_{ \rm IR }^2 }{ g_{ \star }^{ \rm gauge } } \rho^{ \mu }_{ \rm IR }
%\eea
%

%
%-- 5D (for example, simply evaluating usual gauge KK profile at the Higgs brane, assuming no localized kinetic terms for
%gauge fields there) gives
%
%\bea
%c & = & ...
%\eea
%

\noindent {\bf Electroweak and flavor/CP violation precision tests}

\begin{figure}
\centering
\includegraphics[height=55mm]{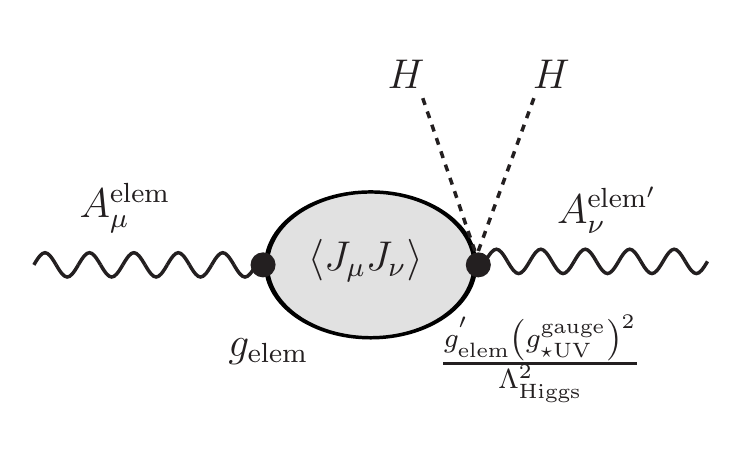}
\caption{
%Process that contributes to S-parameter of EW precision test
%
Contribution to the S-parameter 
from the IR strong dynamics.
%
%This involves one universal (denoted as $g_{\rm univ}$) and one non-universal ($g_{\rm non-univ}$) couplings. This diagram is obtained by ``sewing'' $A_{\mu}^{\rm elem} J^{\mu}_{\rm strong \; IR}$ (universal) and $J^{\nu}_{\rm strong \; IR} \frac{J_{\nu}^{t/H}}{\Lambda_{\rm Higgs}^2}$ (non-universal) together and treating $\frac{J_{\nu}^{t/H}}{\Lambda_{\rm Higgs}^2}$ as a ``background'' gauge field. \sh{Please check and see if it looks okay.}
%
%
}
\label{fig:S_parameter}
\end{figure}

The physics of top/Higgs compositeness with characteristic mass scale $\sim \Lambda_{\rm Higgs}$ (where the UV strong dynamics confines) contributes to EW and flavor/CP violation precision tests.
%
%Composite resonances of UV CFT with mass $\sim \Lambda_{\rm Higgs}$ contribute to EW and flavor/CP violation precision tests. 
%
However, as we already indicated at the beginning of section \ref{3brane}, these contributions are safe from experimental constraints for the choice of $\Lambda_{\rm Higgs} \sim \mathcal{O} (10)$ TeV.
Notice that the (small) flavor non-universal parts of the couplings of spin-1 resonances of the {\em IR} 
strong dynamics [see Eq.~(\ref{gauge_topHiggs})] -- which are suppressed by $\sim \Lambda_{ \rm Higgs }$ -- {\em also} give contributions (via their virtual exchange) to EW and flavor/CP violation precision tests.
However, as we will show now, such effects are comparable to the {\em direct} (albeit still virtual) 
effects of $\Lambda_{ \rm Higgs }$ scale physics 
%
%exchange of resonances of UV CFT of mass $\Lambda_{ \rm Higgs }$, 
%
hence are safe/on the edge (just like the latter). %\footnote{In this discussion, mostly for the simplicity of discussion, instead of {\em hadron} language of 4D strong dynamics, we will use {\em operator} language. We also performed the computation using hadronic picture and 5D dual and confirmed the results presented here. \sh{Is it okay to have this footnote ? We can remove if that's better.}}

We begin our discussion by considering contributions of IR strong dynamics to 
precision tests observables using the above non-universal coupling only {\em once}, for example, 
the operator corresponding to the
$S$-parameter:
\bea
\delta {\cal L } & \sim & C W_{ \mu \nu }^3 B_{ \mu \nu } H^{ \dagger } H, \; \rm with \nonumber \\
C & \equiv &  \frac{ g g^{ \prime } S }{ 16 \pi v^2 },
\eea
$W_3$ and $B$ being the neutral $SU(2)$ and hypercharge gauge fields and $g$ ($g'$) are the respective 
%
%$SU(2)_{\rm L}$($U(1)$) 
%
gauge couplings.
Integrating out physics at and above the scale $\sim \Lambda_{\rm Higgs}$ generates in the IR effective theory the above operator with coefficient $C_{ \rm UV } \sim g g^{ \prime } / \Lambda_{ \rm Higgs }^2$ (based on usual, naive dimensional analysis).
The contribution from the IR strong dynamics can be obtained by computing the diagram shown in Fig.~\ref{fig:S_parameter}. Such a diagram can be generated by sewing together Eq.~(\ref{eq:lagrangian_J_CFT IR_to_t/H}) (non-universal coupling) and the (universal) coupling $A_{ \mu } J^{ \mu }_{ \rm strong \; IR }$ [mentioned below Eq.~(\ref{lagrangian_rho_photon_mixing})], via the common $J^{  \mu }_{ \rm strong \; IR }$.
\bea
g_{ \rm elem } \left( g_{\star \; \rm UV}^{\rm gauge} \right)^2 A_{\mu}^{\rm elem} \langle J^{\mu}_{\rm strong \; IR} J^{\nu}_{\rm strong \; IR} \rangle \frac{J_{\nu}^{\rm t/H}}{\Lambda_{\rm Higgs}^2}.
\label{eq:IR_CFT_to_S-parameter_operator_picture}
\eea
%
%\bea
%A_{\mu}^{\rm elem} \underbrace{\langle J^{\mu}_{\rm CFT \; IR} J^{\nu}_{\rm CFT \; IR} \rangle}_{\sim p^2 \log p^2} \frac{J_{\nu}^{\rm t/H}}{\Lambda_{\rm Higgs}^2}.
%\eea
%
%
%In momentum space, for large Euclidean momentum, $p \gg \Lambda_{\rm IR}$, the $p^2$ dependence of current-current correlator is %determined by conformal invariance and by the fact that $J_{\mu}$ is a conserved current: 
%\bea
%\left\langle 0 \left\vert J^{\mu}_{\rm strong \; IR} (p) J^{\nu}_{\rm strong \; IR} (-p) \right\vert 0 \right\rangle \sim \left( \eta^{\mu\nu} - \frac%{p^{\mu} p^{\nu}}{p^2} \right) p^2 \log p^2. 
%\label{eq:JJ in p-space}
%\eea
%
We then extract two derivatives/momenta
out of the current-current correlator (since only that piece contributes to the $S$-parameter operator) and use naive dimensional analysis, thus finding a log-divergence in the theory below $\Lambda_{ \rm Higgs }$.
Finally, matching to the $S$-parameter operator and using Eq.~(\ref{bCFT}) for overall size of correlator, 
we get $C_{ \rm strong \;  IR } \sim g g^{ \prime } \log \left( \Lambda_{ \rm Higgs } / \Lambda_{ \rm IR } 
\right) / \Lambda_{ \rm Higgs }^2$. 
As already mentioned above, the {\em total} contribution of the IR strong dynamics to $S$-parameter 
%
%in 
%Eq.~(\ref{eq:IR_CFT_to_S-parameter_operator_picture}) 
%
is then comparable to that 
from physics at $\Lambda_{ \rm Higgs }$.
However, there is an important feature we want to emphasize. Namely, the contribution of IR strong dynamics to $S$-parameter shows a mild {\em logarithmic} enhancement! This enhancement, however, is not harmful because, with custodial symmetry protection, the constraint from EW precision test on the Higgs compositeness scale in the minimal model of
Fig.~\ref{fig:OriginalRS} can be as low as $\sim 3$ TeV \cite{Agashe:2003zs} so that, even with the above enhancement in the extension in Fig.~\ref{fig:3brane_w/0_A5}, the overall size is small enough with $\Lambda_{\rm Higgs} \sim O (10)$ TeV.

Next, we consider cases where {\em two} non-universal couplings are involved, giving (for example) a 4-top quark operator, which after rotation to mass basis for quarks will give flavor-violating effects even for light fermions such as $K-\bar{K}$ mixing \cite{Gherghetta:2000qt, Csaki:2008zd}. 
Clearly, the contribution of UV strong dynamics to such effects is $\propto 1 / \Lambda_{ \rm Higgs }^2$ (just like for
$S$-parameter above).
For the IR strong dynamics contribution, we combine Eq.~(\ref{eq:lagrangian_J_CFT IR_to_t/H}) with itself in this case.
Here, the current-current correlator can instead give a {\em quadratic} divergence, which reduces the initial $\sim 1 / \Lambda_{ \rm Higgs }^4$ suppression by two powers. 
%
%\sh{It is not clear to me (i) how we get quadratic-div from current-current correlator of Eq.~\ref{eq:JJ in p-space} (I know we see this from hadron picture) and (ii) initial $\sim 1 / \Lambda_{ \rm Higgs }^4$ from Eq.~\ref{eq:IR_CFT_to_S-parameter_operator_picture}. For (i), should we add ``explicit'' breaking term to $\langle jj \rangle$ ? which, I think, contains quadratic-div. Otherwise, I will see ``logarithmic'' enhancement even for two non-univ case, which is in {\em dis}-agreement with results from hadron picture. For (ii), is this ``remnant'' of hadron picture, or can be directly seen from Eq.~\ref{eq:IR_CFT_to_S-parameter_operator_picture}?} 
%
That is, the contribution from the entire IR strong dynamics to such flavor/CP violating processes are comparable to that of the physics of the UV strong dynamics, hence safe.

We stress that, for a {\em fixed} $\Lambda_{ \rm Higgs }$, the contribution to 
precision tests from IR strong dynamics is (roughly) {\em independent} of $\Lambda_{ \rm IR }$
so that there is no relevant constraint on $\Lambda_{ \rm IR }$ from here; instead  the bound  on $\Lambda_{ \rm IR }$
is dominated by the direct LHC searches
which will be discussed in section \ref{subsubsec:current bounds in flavor-univ limit}.

\subsubsection{Spin-2/Graviton KK}
\label{subsubsec:spin-2/KK graviton}

We denote the 
%
%KK graviton
%
composite spin-2 by $H^{ \mu \nu }$.
In general, $H^{ \mu \nu }$ couples to not only $T_{ \mu \nu }$ of composites, but also
other possible Lorentz structures built out of the latter fields \cite{Panico:2016ary}.
Here, for simplicity and because it dominates in warped 5D effective field theory, we will use (only) $T_{ \mu \nu }$ as a representative structure
(others will anyway give roughly similar size for coupling/amplitude).
%
%In fact, if we do have a weakly-coupled warped extra-dimensional description of the 4D strong dynamics, then $T_{ \mu \nu }$ actually {\em dominates} (the remaining couplings being suppressed by extra factors of 5D cut-off).
If experiments show spin structures other than $T_{\mu\nu}$, it would point to strong dynamics without a good 5D dual.

\noindent {\bf Coupling to SM gauge bosons} 

\begin{figure}
\centering
\includegraphics[height=60mm]{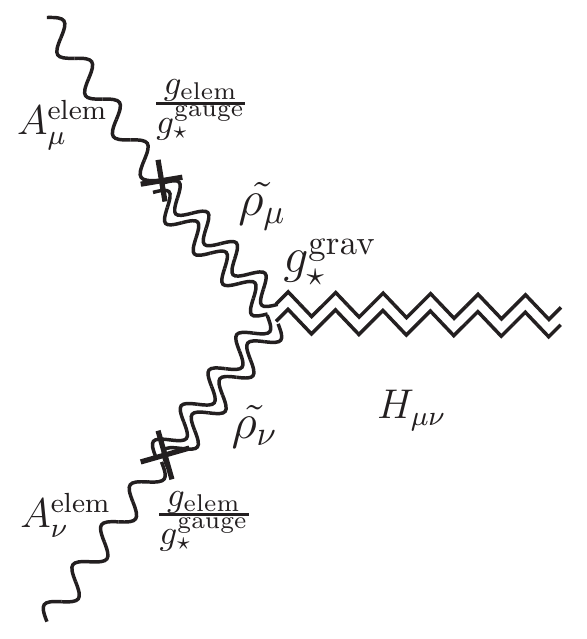}
\caption{Spin-2 KK graviton couplings to SM gauge bosons}
\label{gauge_grav}
\end{figure}

The coupling of $H^{ \mu \nu }$ to {\em SM} gauge bosons
is obtained (see Fig.~\ref{gauge_grav})
by first coupling it to $\tilde{ \rho }$'s with strength $g^{ \rm grav }_{ \star \; \rm IR }$ (i.e., a 3-composite vertex), 
followed by mixing of $\tilde{ \rho }$'s with external gauge field (as outlined above), i.e., 
\bea
\delta {\cal L} \left( \Lambda_{ \rm IR } \right) & \sim &
\frac{ g_{ \star \; \rm IR}^{ \rm grav } }{ \Lambda_{ \rm IR } } 
 H^{ \mu \nu } T_{ \mu \nu }^{ \rm  (\tilde{ \rho }) } \nonumber \\
 & \rightarrow & 
 \left( \frac{ g_{ \rm SM } }{ g_{ \star \; \rm IR}^{ \rm gauge } } \right)^2 \frac{ g_{ \star \; \rm IR}^{ \rm grav } }{ \Lambda_{ \rm IR } } 
 H^{ \mu \nu } T_{ \mu \nu }^{ \rm (gauge) }.
\label{spin-2_gauge}
\eea
%

%-- 5D gives
%
%\bea
%d & = & ...
%\eea
%

%

\noindent {\bf Coupling to radion/dilaton} 

In addition, we have the coupling to two dilatons/radions:
\bea
\delta {\cal L} \left( \Lambda_{ \rm IR } \right) \sim
\frac{ g_{ \star \; \rm IR}^{ \rm grav } }{ \Lambda_{ \rm IR } } 
H^{ \mu \nu } T_{ \mu \nu }^{ (\varphi) }.
\label{spin-2_dilaton}
\eea
Of course, this is relevant for decay of composite spin-2/KK graviton only if $m_{ \varphi } \lesssim \Lambda_{ \rm IR } / 2$ and in this case,
dominates over other decays.

\noindent {\bf Flavor {\em non}-universal coupling (to top/Higgs)} 

Finally, coupling to top quark/Higgs
follows from a procedure similar to spin-1 above, i.e., 
we have
\bea
\delta {\cal L} \left( \Lambda_{ \rm Higgs } \right)
\sim 
\frac{ \left( g_{ \star \rm \; UV}^{ \rm grav} \right)^2 }{ \Lambda_{ \rm Higgs }^4 }
T^{ \mu \nu \; (t/H) }
T_{ \mu \nu }^{ \rm (strong \;  IR) }
\label{twoTs}
\eea
where $T^{ \mu \nu \; (t/H) } (T_{ \mu \nu }^{ \rm (strong \;  IR) })$ is energy-momentum tensor made of top/Higgs fields (preons of IR strong dynamics) and this coupling of top/Higgs to IR strong dynamics is generated by integrating out physics at the scale $\sim \Lambda_{ \rm Higgs }$, with a coupling characteristic of gravity sector of the {\em UV} strong dynamics. %\sh{The following sentence may need to be changed/corrected} We have also assumed that in IR composite theory there is no symmetric tensor operator $\mathcal{O}_{\mu\nu}$ with scaling dimension less than that of $T_{\mu\nu}^{\rm (strong \; IR)}$ which interpolates $H_{\mu\nu}$.
After IR theory hadronizes, Eq.~(\ref{twoTs}) becomes 
\bea
\delta {\cal L} \left( \Lambda_{ \rm IR } \right)
& \sim & 
\left( \frac{ \Lambda_{ \rm IR } }{ \Lambda_{ \rm Higgs } } \right)^4  \frac{ \left( g^{ \rm grav }_{ \star \rm \; UV} \right)^2 }{ g^{ \rm grav }_{ \star \rm \; IR } }  \frac{ H^{ \mu \nu } }{ \Lambda_{ \rm IR } } T_{ \mu \nu }^{ (t/H) } 
\label{graviton_topHiggs}
\eea
using the interpolation 
\bea
T_{ \mu \nu }^{ \rm strong \; IR } 
%
%| 0 \rangle 
%
& \sim & \frac{ \Lambda_{ \rm IR }^3 }{ g_{ \star \rm \; IR}^{ \rm grav} } H_{ \mu \nu }.
\eea
%
%
%-- 5D (again, evaluating the KK graviton profile at the Higgs brane, but here, we have to include modification of KK graviton profile due %to tension on Higgs brane/change of AdS curvature across it) gives
%
%\bea
%e & = & ...
%\eea
%
Production of composite spin-2/KK graviton occurs via coupling to gluons in Eq.~(\ref{spin-2_gauge}). Decays of composite spin-2/KK graviton is dominated by the same couplings, i.e., into {\em all} SM gauge bosons  {\em and} to pair of dilatons via
Eq.~(\ref{spin-2_dilaton}), assuming $m_{ \varphi } < \Lambda_{ \rm IR } / 2$.
We give a summary of relevant couplings in table \ref{table_coupling_3brane}.
Given the above flavor non-universal couplings of KK graviton of the IR strong dynamics
(cf.~those of gauge KK discussed earlier), it is clear that
contributions from KK graviton exchange to precision tests
are suppressed compared to those of gauge KK by $\sim E^2 / \Lambda_{ \rm Higgs }^2$, where $E$ is the characteristic (low) energy of the corresponding test.
Hence, there is no additional constraint here from the KK graviton sector.

%%%%%%%%%%%%%%%%%%%%%%%%%%%%%%%%%%%%%%%
% TABLE
%%%%%%%%%%%%%%%%%%%%%%%%%%%%%%%%%%%%%%%
\begin{table}[tbp]
\begin{adjustwidth}{-1.3cm}{}
\begin{tabular}{ | c || c | c | }
\hline 

\backslashbox[45mm]{Resonance}{Type of coupling}
&\makebox[8em]{Higgs compositeness-sensitive}&\makebox[8em]{flavor-blind}\\ 
\hline\hline

radion/dilaton & $\frac{ g^{ \rm grav }_{ \star \; \rm IR } }{ \Lambda_{ \rm IR } }
\left( \frac{ \Lambda_{ \rm IR } }{ \Lambda_{ \rm Higgs } } \right)^{4-\epsilon} 
\Big[ m_t \bar{t} t + \left( \partial H \right)^2 \Big]$
& $ \frac{ g^{ \rm grav }_{ \star \; \rm IR } }{ \Lambda_{ \rm IR } } \left( \frac{ g_{ \rm SM } }{ g^{ \rm gauge }_{ \star \; \rm IR} } \right)^2 
F^{ \mu \nu } F_{ \mu \nu }$ 

\tabularnewline \hline

KK $Z$ & $\frac{\left(g^{ \rm gauge }_{ \star \; \rm UV }\right)^2}{g^{ \rm gauge }_{ \star \; \rm IR }} \left( \frac{ \Lambda_{ \rm IR } }{ \Lambda_{ \rm Higgs } } \right)^2 \left( \bar{t} \gamma^{ \mu } t
+ H^{\dagger } \partial^{ \mu } H 
\right)$ & $\frac{ g_{ \rm EW }^2 }{ g^{ \rm gauge }_{ \star \; \rm IR} } \left( \bar{q} \gamma^{ \mu } q + \bar{l} \gamma^{ \mu } l \right)$ 
({\em all} generations) 
\tabularnewline \hline

KK gluon & $\frac{\left(g^{ \rm gauge }_{ \star \; \rm UV }\right)^2}{g^{ \rm gauge }_{ \star \; \rm IR }} \left( \frac{ \Lambda_{ \rm IR } }{ \Lambda_{ \rm Higgs } } \right)^2 \bar{t} \gamma^{ \mu } t$
& $\frac{ g_{ \rm QCD }^2 }{ g^{ \rm gauge }_{ \star \; \rm IR} } \bar{q} \gamma^{ \mu } q$ 
({\em all} generations) 
\tabularnewline \hline

KK graviton & $\frac{\left(g^{ \rm grav }_{ \star \; \rm UV }\right)^2}{g^{ \rm grav }_{ \star \; \rm IR }} \frac{ 1 }{ \Lambda_{ \rm IR } } \left( \frac{ \Lambda_{ \rm IR } }{ \Lambda_{ \rm Higgs } } \right)^4  T_{ \mu \nu }^{(t/H)}$
& $\frac{ g^{ \rm grav }_{ \star \; \rm IR} }{ \Lambda_{ \rm IR } } \left( \frac{ g_{ \rm SM } }{ g^{ \rm gauge }_{ \star \; \rm IR} } \right)^2 T_{ \mu \nu }^{\rm (gauge)}$
\tabularnewline
\hline
\end{tabular}
 
\caption{Summary of universal and non-universal couplings of various composites for the model with one intermediate brane. $T_{ \mu \nu }^{(t/H)} (T_{ \mu \nu }^{\rm (gauge)})$ is energy-momentum tensor made of top/Higgs (SM gauge bosons) fields.}
\label{table_coupling_3brane}
\end{adjustwidth}
\end{table}

\subsection{Extended Bulk Gauge Symmetries/Dual to PNGBs of Vector-Like Confinement}
\label{section:A_5}

{\bf Relation to vector-like confinement}

\begin{figure}

\center 

\includegraphics[width=0.8\linewidth]{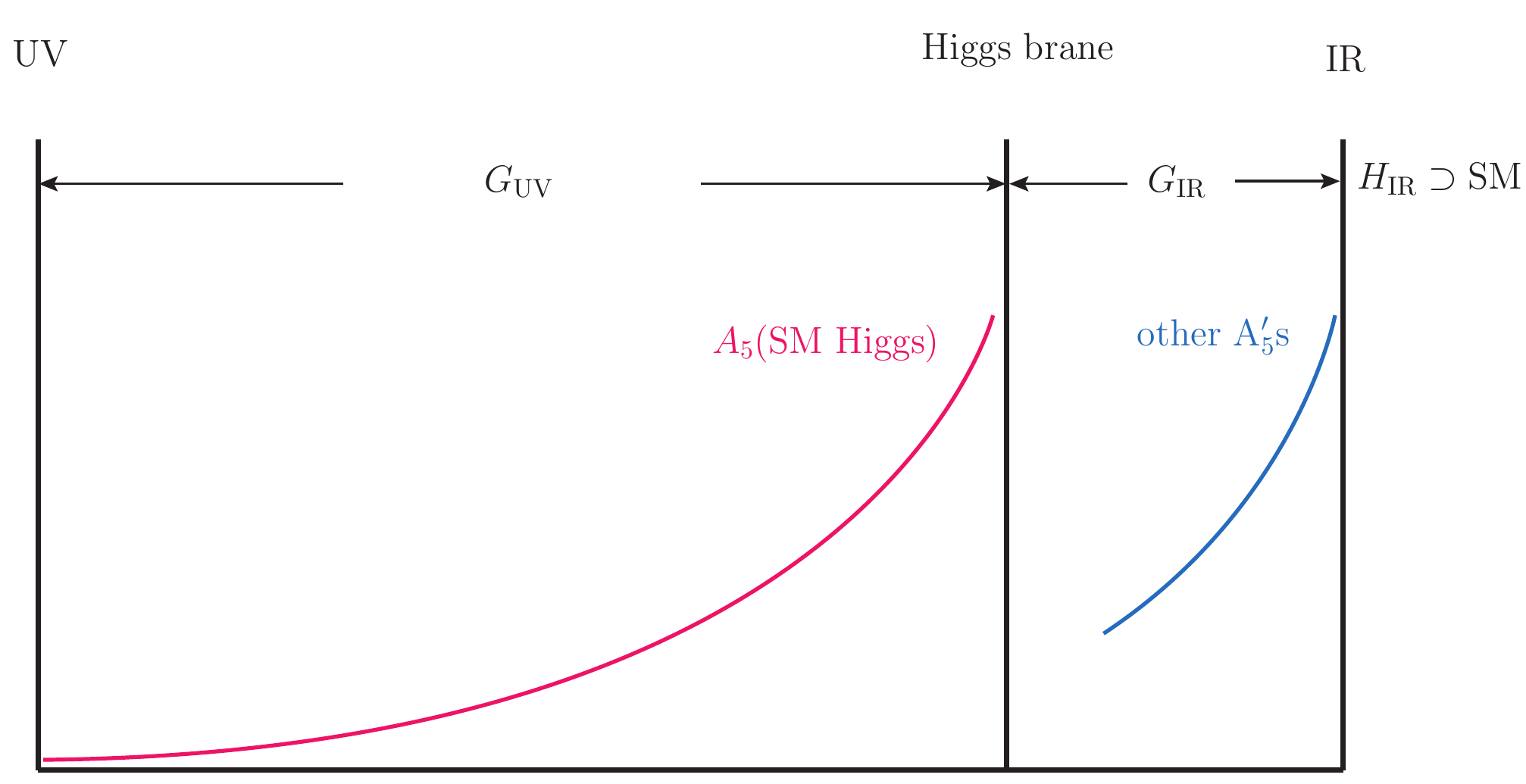}

\caption{Extended bulk gauge symmetries, with rightmost bulk segment being dual to vector-like confinement}
\label{setup_VC}
\end{figure}

We can enlarge the bulk gauge symmetries beyond the SM. We then consider
breaking them down to the smaller groups (while preserving the SM subgroup of course)
on the various branes by simply imposing Dirichlet boundary condition, i.e., 
\bea
G_{ \rm UV } & \stackrel{ \Lambda_{ \rm Higgs } }{ \rightarrow } G_{ \rm IR } & \stackrel{ \Lambda_{ \rm IR } }{ \rightarrow } H _{ \rm IR } 
\supset {\rm SM }
\label{eq:Vector_like_confinement_gauge_group}
\eea
where each stage of gauge symmetry breaking delivers (scalar) $A_5$'s, localized at the corresponding brane (including possibly the
SM Higgs boson in the first step). Such a framework is shown in Figs.~\ref{fig:3brane} and \ref{setup_VC}.
These $A_5$'s are dual to PNGBs arising from spontaneous breakdown of
global symmetries of the strong dynamics corresponding to the gauged ones shown in 
Eq.~(\ref{eq:Vector_like_confinement_gauge_group})
\cite{Contino:2003ve}.
In particular, the 4D physics dual to the last stage of breaking (rightmost bulk in Figs.~\ref{fig:3brane} and \ref{setup_VC}), i.e., SM symmetries being unbroken, 
is known in the literature as vector-like confinement \cite{Kilic:2008pm}.
While from the 5D viewpoint, presence of $A_5$'s seems rather ``non-minimal'', it is quite natural to have
PNGB's in 4D strong dynamics 
as illustrated by ordinary QCD.
In fact, QCD-like strong dynamics was first used to realize the general idea of vector-like confinement.

Note that $A_5$'s are massless at tree-level (in the presence of only the above boundary condition breaking), acquiring a potential via loops, with mass scale being set by corresponding $\Lambda$. 
Thus they are naturally light, as expected from them being dual to PNGB's.
Gauge and graviton KK modes (and even possibly the radion) can then decay into pairs of $A_5$'s, drastically modifying the LHC signals 
of the gauge and graviton KK (or radion) based only on the couplings shown earlier. 
In this paper, we take the {\em minimal} 5D perspective in assuming that $A_5$'s are absent, cf.~the expectation based on
QCD-like 4D strong dynamics.
Hence, gauge KK will decay dominantly into pairs of SM fermions, while SM gauge bosons
will be the search channel for KK graviton and radion, as mentioned earlier.
Remarkably, the 
flexibility afforded by 5D leads to broader class of models, with more diverse phenomenology 
than contemplating just 4D QCD-like strong dynamics.

\noindent {\bf Coupling to two SM gauge bosons}

There is an interesting comparison with dilaton/radion that we would like to draw by
considering the simplest mechanism for production and decay of (single) $A_5$ (dual to PNGB).
Namely, PNGB famously has a coupling to two weakly-coupled gauge bosons via the (gauged)
Wess-Zumino-Witten term, for example,
we have $\pi^0 F_{ \mu \nu } \tilde{F}^{ \mu \nu }$ leading to the decay
$\pi^0 \rightarrow \gamma \gamma$ in real-world QCD.
This interaction is dual to the one originating for the $A_5$ from the Chern-Simons term in the 5D model (see also discussions in \cite{Franceschini:2015kwy, Flacke:2006ad}):
\bea
{\cal L_{ \rm 5D } } & \ni & K f_{ a b c } \epsilon^{ MNRST } A^a_M F^b_{ NR } F^c_{ ST } + ...\nonumber \\
& \sim & 
K f_{ a b c }  A^a_5 F^{ b \; \mu \nu } \tilde{F}^c_{ \mu \nu } +...
\label{CS}
\eea
where $a$, $b$, $c$ are gauge adjoint indices.

Crucially, we see that, {\em ir}respective of considerations of parity as a  fundamental symmetry, 
the coupling of $A_5$ to two SM gauge bosons via Chern-Simons term
has ``CP-odd'' structure, i.e., involves $F_{ \mu \nu } \tilde{F}^{ \mu \nu }$. This feature is in {\em contrast} to the ``CP-even'' coupling, i.e., to $F_{ \mu \nu } F^{ \mu \nu }$, of dilaton/radion as we see in Eq.~(\ref{coupling_dilaton_gauge}).
Let us compare to vector-like confinement, in particular, QCD-like dynamics: this theory respects parity even in the IR and 
PNGB's are
parity-odd (as per the Vafa-Witten theorem \cite{Vafa:1984xg}), which enforces a coupling to pairs of SM gauge bosons to be to the combination $F_{ \mu \nu } \tilde{F}^{ \mu \nu }$. However, we see that there is a more 
general (than parity) argument for such a structure from Chern-Simons term in 5D.

Moreover, the 5D Chern-Simons term is {\em dual} to anomalies in global currents of the 4D strong dynamics,
i.e., $K$ of Eq.~(\ref{CS}) -- appropriately made dimensionless -- 
is related to the coefficient of the  {\em chiral} anomaly in 4D. %(which can be computed for certain QCD-like theories). 
In this sense, we see that there is actually a {\em similarity} in the couplings of $A_5$ (PNGB) and dilaton to
two SM gauge bosons, i.e., both are driven by anomalies: chiral for former vs.~{\em scale} anomaly for dilaton [as seen clearly in
first line of Eq.~(\ref{coupling_dilaton_gauge}), i.e., the ``$b_{ \rm strong }$''-form].

\section{Phenomenology}

\label{pheno}

%
%-- \ka{I suggest we should {\em not} focus too much on disambiguation from ``impostors'', especially now that
%750 GeV might be history. Instead, we could just highlight distinctive features of our framework, which
%will anyway, i.e., {\em in}directly, suggest how to distinguish from others...}
%

{\bf General features} 

We first discuss some overall points, before studying each particle in detail.
%
%
%Needless to say, 
%
Assuming $\Lambda_{ \rm Higgs } \gg \Lambda_{ \rm IR }$, the couplings of 
the (lightest) KK/composite spin-1 gauge bosons to the SM matter (fermions and Higgs) are {\em significant} (albeit mildly 
suppressed
relative to the SM values)
and (approximately)
flavor-{\em blind}: see last term in Eq.~(\ref{coupling_rho}) and Eq.~(\ref{gauge_topHiggs}).
On the other hand, 
radion and KK/composite graviton couple predominantly to pairs of SM gauge bosons and {\em negligibly} to SM 
matter: see Eqs.~(\ref{spin-2_gauge}), (\ref{graviton_topHiggs}), (\ref{coupling_dilaton_gauge}) and (\ref{radion_topHiggs}).
This feature is in sharp contrast to {\em standard} minimal model of Fig.~\ref{fig:OriginalRS}, where couplings to heavy SM (top quark/Higgs/longitudinal
W$/Z$) dominate as far as decays are concerned. So, dilepton, diphoton and dijet final states are usually -- and correctly -- 
{\em neglected}, but now they acquire significance or even the dominant role.
At the same time, the (small) flavor non-universality arising in these couplings (i.e., Higgs/top compositeness scale) can be probed 
by {\em precision} studies of these resonances (of mass $\sim \Lambda_{ \rm IR }$),
thereby distinguishing it from (purely) vector-like confinement (which corresponds to decoupling of top/Higgs
compositeness scale), rather
experimentally one can see the latter as a vestige of a full solution to the Planck-weak hierarchy.
Finally, in the case of a unified bulk gauge symmetry, i.e., {\em entire} SM gauge group is subgroup of {\em simple} IR bulk gauge group ($ H_{ \rm IR }$ of Eq.~(\ref{eq:Vector_like_confinement_gauge_group})), we should of course also find that resonances come in complete degenerate unified multiplets. This is dual to the IR strong dynamics having a simple global symmetry partially gauged by SM.

\subsection{Radion/dilaton}

\noindent
{\bf Dilaton Production}

Note that dilaton can be somewhat lighter than higher spin composites [see Eq.~(\ref{radion_mass})], thus possibly the first particle to be discovered.
Rough estimates of the (total) cross-section (from gluon fusion) for $g^{ \rm grav }_{ \star } = 3$;
$g^{ \rm QCD}_{ \star } = 3$, $\Lambda_{ \rm IR } = 3$ TeV and $m_{ \varphi } = $1 (2)  TeV are $\sigma_{\rm tot} \sim 600$ ($\sim 50$) fb.\footnote{All cross-section numbers are obtained using implementations of above models into Madgraph.} 
One of these sample points was mentioned as part of table \ref{table_radion_KK gravity} in introduction. 

\noindent
{\bf Dilaton Decay}

Moving onto decays of dilaton, these are dominantly to two SM gauge bosons
(based on the couplings discussed earlier, assuming $\Lambda_{ \rm Higgs } \gg \Lambda_{ \rm IR }$).
It is noteworthy that in the unified case, i.e.,
SM gauges a subgroup of a simple global symmetry group of 4D strong dynamics
(so that $g^{ \rm gauge }_{ \star }$ or $b_{ \rm strong }$ is 
{\em universal}), we get [see Eq.~(\ref{coupling_dilaton_gauge})]
\bea
\frac{ \hbox{dilaton coupling to two SM gauge bosons} } { \left(\hbox{
%
%respective SM 
%
gauge coupling}\right)^2 } & \approx & \hbox{same 
%
%universal 
%
for {\em all} SM gauge groups}.
\label{univ_dilaton}
\eea
%
%In particular, in the case of universal $g^{ \rm gauge }_{ \star }$, the 
%prediction in Eq.~(\ref{univ_dilaton}) becomes decay width $\propto$ coupling of radion$^2$ $\propto $ SM gauge coupling$^4$
%(of course, up to degrees of freedom counting); 
%
This striking feature can be checked by measuring dilaton BR's.
%once more than one decay widths ($\propto$ coupling of radion$^2$, up to degrees of freedom counting)  is measured, for example, $\gamma \gamma$ and $WW$ or $ZZ$, in case gg is more difficult to do.
%
Numerically, BR's to $\gamma \gamma$, $ZZ$, $WW$ and $gg$ are $\approx$ 0.1 $\%$, $\lesssim$ 4$\%$, 4$\%$ and 90$\%$, respectively.
However, note that the above universality (among the SM gauge groups) feature applies for {\em any} $H_{\rm IR}$-singlet composite scalar. In this sense dilaton is not unique.
The current bounds on cross-section $\times$ BR to di-photons from resonant di-photon searches at the LHC \cite{Aaboud:2016tru}
are $\sim$ a few (1) fb for 1 (2) TeV mass.
Similarly, di-jet searches \cite{Khachatryan:2015sja} give a bound of $1 / $( a few) pb 
for 2 TeV mass, but none for 1 TeV. %(presumably because background increases rapidly at low masses).
Both of these are 
satisfied for the above illustrative choice of parameters.

\noindent
{\bf CP structure}

The CP-even structure of the couplings to SM gauge bosons for dilaton vs.~CP-odd for $A_5$/PNGB's (discussed above: see Eqs.~(\ref{CS}) and (\ref{coupling_dilaton_gauge})) is an important issue. It can be discriminated by
(for example) decays to $ZZ \rightarrow$ four leptons, using the additional observables therein, i.e., corresponding to polarization of $Z$
(as compared to using just angular distribution of spin-summed SM gauge boson taken as ``final'' state, which is the same for both cases) \cite{Stolarski:2012ps}.

\subsection{Spin-1 composite}

Here, we have more than one type, each with several competing decay channels. So, we need more detailed analysis
%
%estimates
%
for obtaining bounds/signals. 
We give some general arguments first.
In the {\em unified} case, based on same mass and composite coupling as in Eq.~(\ref{coupling_rho}), we should find
(similarly to the radion above)
\bea
\frac{ 
\hbox{composite spin-1
%
%production 
%
{\em cross-section}} }{ \left(\hbox{
%
%respective 
%
%corresponding 
%
%SM 
%
SM gauge coupling}\right)^4 } & \approx &  
\hbox{same for {\em all} SM gauge groups.}
%
%universal
%
\label{univ_spin1}
\eea
In the {\em non}-unified case, while the above relations do not apply, interestingly
the following {\em correlation} between radion decays and spin-1 production cross-section can then
be tested: as seen from Eqs.~(\ref{coupling_dilaton_gauge}) and (\ref{coupling_rho}), we expect 
\bea
\frac{ 
\hbox{coupling of dilaton to SM gauge boson} \times 
\hbox{(gauge 
%
%SM 
%
coupling)}^2 
}
{ 
\hbox{(corresponding) composite spin-1
%
%production 
%
{\em cross-section}} }
& \approx & 
%
%\rm universal
%
\hbox{same for {\em all} SM gauge groups}
\nonumber \\
& \propto & g^{ \rm grav }_{ \star }
\label{univ_spin1_dilaton}
\eea
i.e., remarkably, in spite of apparent  lack of unification (i.e., $b_{ \rm strong}$ is different for different gauge groups),
we find that the above ratio {\em is} universal! Moreover, it applies only for the case of composite scalar being dilaton, i.e.,
the above relation is {\em not} valid for a generic scalar composite.
In contrast, in the unified case, the above correlation is {\em not} independent of the
two separate relations discussed earlier, i.e., Eqs.~(\ref{univ_dilaton}) and (\ref{univ_spin1}).
Note that the universal constant on RHS of Eq.~(\ref{univ_spin1_dilaton}) involves 
$g^{ \rm grav }_{ \star }$ [apart from other known factors: see Eq.~(\ref{coupling_dilaton_gauge}) and last term of Eq.~(\ref{coupling_rho})]. 
Thus,  
independent determination of 
$g^{ \rm grav }_{ \star }$, for example, from KK graviton measurements could provide an
interesting test of this framework 
using Eq.~(\ref{univ_spin1_dilaton}).
This would apply to {\em both} unified and non-unified cases discussed above.

\subsubsection{Current bounds in flavor-universal limit}
\label{subsubsec:current bounds in flavor-univ limit}

Based on the suppressed (as compared to the SM, but still non-negligible) and flavor-universal coupling in the last term of Eq.~(\ref{coupling_rho}), 
we find that 
spin-1 masses of a {\em few} TeV are still consistent with the LHC searches performed
so far in {\em multiple} channels .
We now move onto more details, discussing bounds on KK $Z$ first, followed by KK gluon.

\noindent
{\bf KK $Z$}

\noindent (i) Di-lepton:

Note that composite/KK $Z$ in this case is (approximately) like sequential SM $Z^{ \prime }$, but with coupling to light quarks
inside proton (the dominant production mechanism) being {\em reduced} by
$\sim g_{ \rm EW } / g^{ \rm EW }_{ \star }$.
We find that predicted cross-section of sequential SM $Z^{ \prime }$ exceeds the 
%
%run 1 
%
bound
%
%\footnote{2015 data gives roughly similar, but just {\em slightly} weaker bound \cite{CMS:2015nhc}.} 
%
\cite{Khachatryan:2014fba} by $\sim 80 \; (20)$ for $M_{ Z^{ \prime } } \sim 1.5  \; (2)$ TeV.
Translating this bound to our case, we get (setting $g_{ \rm EW } \sim 0.7$):
\bea
\Lambda_{ \rm IR } & \gtrsim &  1.5 \; \hbox{TeV}  \; \hbox{for} \;  g^{ \rm EW }_{ \star } \sim 6 \\
 & \gtrsim &  2.0 \; \hbox{TeV} \; \hbox{for} \;  g^{ \rm EW }_{ \star } \sim 3 
\eea
Of course, only the smaller values of $g^{ \rm gauge }_{ \star } ( \sim 3)$ are 
compatible with a controlled 5D description,
but the somewhat larger values ($\sim 6$) are still reasonable from the viewpoint of 
(purely) 4D strong dynamics,
for example, $\rho \pi \pi$ coupling in real-world QCD is roughly of this size.
We can of course interpolate for other composite spin-1 masses.
To be more precise, we will have to add bound from composite photon (above was just composite $Z$) but as an estimate what we did should suffice.
Also, we have checked that bounds from composite/KK $W$ are somewhat weaker, 
%
%presumably 
%
since 
in the leptonic channel, it is {\em not} quite a resonant search.

\noindent (ii) Di-boson:

%On the other hand, {\em un}like sequential $Z^{ \prime }$, 
Even in the flavor-universal limit ($\Lambda_{\rm Higgs} \to \infty$) KK $Z/W$ couples also to Higgs (including longitudinal $W/Z$, i.e., ``di-bosons'').
So, we can rescale from bound for heavy vector triplet (HVT) model \cite{Pappadopulo:2014qza},
which is (roughly) similar to 
standard warped/composite case of Fig.~\ref{fig:OriginalRS} (i.e., couplings to Higgs/top dominate):
%
%ATLAS 
%
The current bound 
\cite{Aaboud:2016okv} on the mass is 2.6 TeV for $g^{ \rm EW }_{ \star } = 3$.
%
%whereas CMS \cite{CMS:2015nmz} has slightly weaker, i.e., $\sim 2$ TeV (presumably because they combined less number of %channels).
%
However, composite $W/Z$ decay for above  HVT model is $\approx 100 \%$ to dibosons, since
couplings to dibosons are (much) {\em larger} than to the SM fermions, latter being assumed to be flavor-universal.
On the other hand, 
in the (fully) flavor-universal limit that we are considering here, we can readily estimate that it is reduced to (roughly) 10\%, in which case, bound is
weaker than 2 TeV (rescaling from their plots).
%
%even for ATLAS.
%

So, we conclude that di-lepton bound for our KK $Z$ case is a bit stronger than di-boson.

Just for completeness' sake, we mention that there is also a $Z^{ \prime }$ bound of 2 TeV 
%
%CMS
%
from the di-jet search \cite{Khachatryan:2015sja}. However,  
%
%presumably 
%
this assumes coupling to light quarks inside proton is same as SM $Z$, vs.~smaller here.
Similarly, $Z^{ \prime }$ bound from di-top is $\sim 2.5$ TeV 
%
%CMS%
%
\cite{Khachatryan:2015sma}, but that 
%
%seems 
%
is
for a model
with {\em enhanced} (even with respect to the SM) coupling to first {\em and} third generations \cite{Harris:2011ez}; hence for our case, bound should be weaker.
Overall, then di-jet and d-top bounds for KK $Z$ are sub-dominant to that from di-lepton discussed earlier.

\noindent {\bf KK gluon}

\noindent (i) Di-top:

Similarly to KK $Z/W$ above, we can 
rescale from 
%
%run 1 
%
the KK gluon bounds
%
%\footnote{2015 data gives roughly similar -- actually just {\em slightly} stronger -- bound 
%\cite{CMS:2016zte}.} 
%
\cite{Khachatryan:2015sma}:
the
predicted cross-section [all for $g_{ \star }^{ \rm QCD } \sim 5$, as assumed in \cite{Agashe:2006hk}, which is 
quoted in \cite{Khachatryan:2015sma}] is
larger than bound by $\sim$ 6 (2) for mass of KK gluon of 2 (2.5) TeV.
The above bounds are assuming BR to top quarks $\approx 1$ (as in the standard scenario) so that for our case (i.e., with BR to top quarks of $\approx 1/6$ instead), we get
\bea
\Lambda_{ \rm IR } &   \gtrsim  &  2.0 \; \hbox{TeV} \; \hbox{for} \; g^{ \rm QCD }_{ \star } \sim 5  \\
 & \gtrsim & 2.5 \; \hbox{TeV} \; \hbox{for} \;  g^{ \rm QCD }_{ \star } \sim  3
\eea
As usual, we can interpolate for other composite spin-1 masses.

\noindent (ii) Di-jet:

Here, we can re-scale from axigluon 
%
%(or flavor universal coloron) 
%
bounds \cite{Khachatryan:2015sja}, i.e., coupling to our composite gluon  
is smaller by a factor of 
$\sim g_{ \rm QCD } / \left( g^{ \rm QCD }_{ \star } \times \sqrt{2} \right)$,
since coupling of axigluon [see discussion in \cite{Chivukula:2013xla} referred to by 
%
%CMS
%
\cite{Khachatryan:2015sja}] 
is larger than QCD by $\sqrt{2}$.
%
%The 
%2015 data
%
%\footnote{run 1 gives roughly similar, but just {\em slightly} weaker bound \cite{Khachatryan:2015sja}.} 
%
%gives 
%\cite{Khachatryan:2015dcf} 
%
%CMS 
%
The cross-section is 
constrained to be smaller than the prediction for axigluon by $\sim 30  \; (20)$ for axigluon mass of
2 (2.5) TeV.
So, using the above couplings, we get for our case:
\bea
\Lambda_{ \rm IR }  & \gtrsim &  2.0 \; \hbox{TeV}  \; \hbox{for} \; g^{ \rm QCD }_{ \star } \sim  4 \\
 & \gtrsim & 2.5 \; \hbox{TeV} \; \hbox{for} \;  g^{ \rm QCD }_{ \star } \sim  3
\eea
Similarly, we can find the bound for other values of $\Lambda_{ \rm IR }$.

So, di-top and di-jet bound are (roughly) {\em comparable} in the case of KK gluon.

\subsubsection{Probing top/Higgs compositeness}

Next, we 
discuss the possibility of 
being able to see some remnants of top/Higgs compositeness in the
properties of composite resonances at $\Lambda_{ \rm IR }$.

\noindent {\bf Summary}
 
As seen from Eqs.~(\ref{gauge_topHiggs}), (\ref{graviton_topHiggs}) and (\ref{radion_topHiggs}), 
%
%{\em on}-shell 
%
spin-1 
%
%decays
%
couplings (cf.~dilaton and spin-2) at the LHC are most sensitive to flavor {\em non}-universal corrections. 
In particular, for spin-1 composite, the net coupling [combining Eqs.~(\ref{coupling_rho}) and (\ref{gauge_topHiggs})] 
to SM fermions is then given (schematically) by:
\bea
\delta {\cal L } & \sim & \Big[ - \frac{ g_{ \rm SM }^2 } { g_{ \star \; \rm IR }^{ \rm gauge } } + h \frac{ g^{ \rm gauge \; 2 }_{ \star \; \rm UV } }
{ g^{ \rm gauge }_{ \star \; \rm IR } } \left( \frac{ \Lambda_{ \rm IR } }{ \Lambda_{ \rm Higgs } } \right)^2
\Big] \bar{q} \gamma^{ \mu } \rho_{ \mu } q 
\label{net_coupling_rho}
\eea
Here, $h$ is an $O(1)$ factor which depends on details of the model (whether 
a 4D composite theory or 5D dual).
Note that the 5D model gives {\em opposite} sign for the flavor {\em non}-universal coupling (to top/Higgs) of spin-1 vs.~flavor universal
one, i.e., $h > 0$, whereas from purely 4D CFT viewpoint, $h < 0$ cannot be ruled out.
%
%{\em positive}
%
Eq.~(\ref{net_coupling_rho}) shows that the non-universal contributions (second term above)
start becoming 
relevant (i.e., {\em comparable} to the universal first term) for:
\bea
\frac{ \Lambda_{ \rm Higgs }}{ \Lambda_{ \rm IR } } & \sim & \frac{ g_{ \star \; \rm UV}^{ \rm gauge} }{ g_{ \rm SM } } 
\label{sensitivity}
\eea
Setting $\Lambda_{ \rm IR } \sim 3$ TeV; a universal $g^{ \rm gauge }_{ \star \; \rm UV } \sim 3$; $g_{ \rm EW } \sim 0.6$ and $g_{ \rm QCD } \sim 1$, we see that above equality occurs (roughly) for
\bea
\Lambda_{ \rm Higgs } & \sim & 10 \; (15) \; \hbox{TeV for KK} \; \hbox{gluon} \;(Z)
\eea
which 
is (roughly) the flavor bound, i.e., (in general) we {\em do} expect sensitivity to top/Higgs compositeness!
Again, note that in the standard scenario, i.e., $\Lambda_{ \rm Higgs } \sim \Lambda_{ \rm IR }$, the
non-universal contribution actually {\em dominates}: see Eq.~(\ref{net_coupling_rho}).

\noindent {\bf KK gluon vs.~KK $Z$}

In particular, KK gluon might be especially promising in this regard, since
for the flavor-universal case, di-jet bounds on KK gluon 
%
%\cite{Khachatryan:2015dcf}
%
seem comparable to di-top
%
%\cite{Khachatryan:2015sma} 
%
as indicated above, which suggests that there should be significant sensitivity to above perturbations, for example, 
non-universal coupling to top being {\em comparable} to universal might then show up even at {\em discovery} stage!
Whereas, in flavor-universal limit, it seems bounds from di-boson/di-top are somewhat weaker than from di-lepton final state
for KK $Z$, thus suggesting that probe of top/Higgs compositeness (again, for the case when flavor non-universal couplings
are comparable to flavor universal ones) might have to wait for {\em post}-discovery precision-level studies.
On the other hand, as discussed above, for the {\em same} top/Higgs compositeness scale, flavor non-universal effects 
are actually a bit {\em larger} for KK $Z$ than for KK gluon. So, overall, the two modes might be complementary in this regard.

%
%\ka{We could show some numbers/plots for KK gluon as per above discussion...}
%

\noindent {\bf Details of analysis}

Estimates of various BR's illustrating the above ideas are given in table \ref{table_KKZ_KKg}: these were already mentioned 
%
%(or alluded to) 
%
in the introduction, including 
%
%showing 
%
the tables.
We now present more details. 
First, as  a reminder, 
in this table \ref{table_KKZ_KKg}, we fix KK $Z$/gluon mass to be 3 TeV and the composite gauge coupling ($g_{ \star \; \rm UV }^{ \rm gauge }$)
to be 3.
Hence, the 
%
%coupling to  light quarks 
%
production
cross-section is the same throughout the tables, but we vary Higgs compositeness 
scale.

These numbers are obtained simply using the 
net coupling given in Eq.~(\ref{net_coupling_rho}).
%
%flavor
%universal coupling in Eq.~(\ref{coupling_rho}) and the non-universal ones to top/Higgs
%of Eq.~(\ref{gauge_topHiggs}).
%
Just for the sake of concreteness,  
we choose a ``central'' value for the $O(1)$ coefficient $h$ in Eq.~(\ref{net_coupling_rho}) 
so that  $\Lambda_{ \rm Higgs } =$ 10 and 15 TeV gives {\em exact} equality between 
the two terms there for KK gluon and KK $Z$, respectively.
Then, for each $\Lambda_{ \rm Higgs }$, we vary $h$ between a factor  of $2$ and $1/2$ around 
this central value.
Thus, we obtain a {\em range} of BR's even for 
{\em fixed} $\Lambda_{ \rm Higgs }$.
Mostly for simplicity, we assume only $t_R$ (and Higgs) is (fully) composite, i.e., $(t,b)_L$'s compositeness is smaller.
Also, 
we will assume of $h > 0$ (based on 5D model, as mentioned above). 
We then see that  
%
%the 5D model gives {\em opposite} sign for the flavor {\em non}-universal coupling (to top/Higgs) of spin-1 vs.~flavor universal
%one. Thus, 
%
for values of $\Lambda_{ \rm Higgs } / \Lambda_{ \rm IR }$ around Eq.~(\ref{sensitivity}), there is actually a possibility of \;``cancellation'' between the two terms in Eq.~(\ref{net_coupling_rho});
this feature is reflected in these tables in BR's to top/dibosons 
becoming {\em smaller} than flavor-universal limit as we start lowering the Higgs compositeness scale from a high value.
Note that, as reflected by our $O(1)$ variation of $h$ factor, we are not really contemplating a fine-tuning here, rather 
only pointing out that a mild suppression is possible in this way.
Eventually, i.e., for even lower $\Lambda_{ \rm Higgs }$, of course the non-universal part of couplings to top/Higgs 
dominates over universal one so that BR's to top/Higgs become larger, as they asymptote to the
values of minimal model of Fig.~\ref{fig:OriginalRS}. 
Finally, we have to consider the decay of (composite) spin-1 to a dilaton and a SM gauge boson.
Based on Eqs.~(\ref{coupling_dilaton_gauge}) and (\ref{dilaton_rho_photon}), it is straightforward to show that 
there exists choices of the relevant parameters such that this decay 
is (much) smaller than to the SM fermions.
For simplicity, here we assume that 
is the case in tables shown above.
%
%this is realized
%[even if couplings per say, i.e., in Eqs.~(\ref{coupling_dilaton_gauge}) and (\ref{dilaton_rho_photon})] are similar],
%for example, for the choice $m_{ \varphi } \lesssim \Lambda_{ \rm IR }$ (say,
%a factor of $ \gtrsim 1/2$) so that {\em derivative}  associated with the former coupling 
%gives a (even if mild) suppression due to 
%the (small) momentum of final state particles in this case, in addition to phase-space factors being small.
%
%In any case, i.e., even if we do not assume such a mass spectrum, 
%
In other regions of parameter space, 
the decay to a dilaton and a SM gauge boson
will at most be comparable to that to SM fermions and so our choice above (i.e., neglecting the decay to dilaton) is 
reasonable 
for the purpose of estimates here.
Having said this, a dilaton and a SM gauge boson is an 
%
%novel/
%
interesting final state (followed by dilation $\rightarrow$ two SM gauge bosons), 
which (to the best of our knowledge) 
has not been studied before; 
we hope to return to an analysis of it 
in the future.

As anticipated earlier (but now seen more 
explicitly in the tables), as we lower Higgs compositeness scale from decoupling limit, at $\sim O(10)$ TeV, 
we start seeing $\sim O(1)$ deviations from flavor-blindness (middle vs.~rightmost columns), that too ``earlier'' for KK $Z$ than for 
KK gluon. At the same time, these BR's significantly different than standard Higgs compositeness case (leftmost column).
So, the moral here is that composite $Z$/gluon can provide ``glimpse'' into Higgs/top compositeness,
provided that this scale is not too far from the lower limit from flavor/CP violation, i.e., $\sim O(10)$ TeV.

\noindent 
{\bf Other values of KK masses}

For the sake of completeness, we mention that the 
(total) cross-sections for 2 and 4 TeV composite/KK $Z$ and gluon for $g^{ \rm gauge }_{ \star \; \rm IR } = 3$
are $\sim$ 50, 0.5 fb ($Z$) and $\sim$ 2000, 100 fb (gluon), respectively (of course, the 2 TeV case might be ruled out as per above discussion,
unless we invoke extra decay modes, for example to light $A_5$'s). 
From Eq.~(\ref{sensitivity}), it is clear that as we vary composite spin-1 masses in this way, 
one could then be sensitive to lower/higher top/Higgs compositeness scale.

\noindent 
{\bf Comparison to other probes
%
%effects 
%
of top/Higgs compositeness}

Let us summarize by comparing the above signals of top/Higgs compositeness scale of $O(10)$ TeV to other 
%
%tests
%
approaches.
One of the standard probes would be existing/upcoming low-energy flavor experiments,
which will be sensitive to $\Lambda_{ \rm Higgs } \sim O(10)$ almost by construction, since $O(10)$ TeV
was chosen to barely satisfy the {\em current} flavor/CP violation bounds.
Of course, this would provide the most indirect view, for example, even if we see a signal, 
we cannot be sure about which underlying new physics it corresponds to, i.e., whether it is
$\Lambda_{ \rm Higgs } \sim O(10)$ TeV of the warped/composite Higgs framework or some thing else.
On the other hand, the most direct signal is possible at a {\em future}
%
%the proposed 
%
100 TeV hadron
collider, where the associated, i.e., $O(10)$ TeV, physics of compositeness can be produced without any suppression. 
%
%{\em on}-shell
%
%
In fact, this could serve as a motivation to build such a machine.

Here, we showed how extending the usual, minimal framework to include a intermediate brane (Fig.~\ref{fig:Multibrane})
results in novel probe of the general framework.
%
%top/Higgs compositeness.
%
Namely, it 
creates a {\em new}
threshold, i.e., a few TeV resonances intermediate in mass between $O(10)$ flavor scale and
the SM/weak scale itself,
whose leading couplings are flavor-universal, rendering such a mass scale
safe from flavor bounds.
This angle actually combines some of the virtues of both the above approaches, for example,
we can {\em directly} produce the relevant particles
%
%states 
%
at the {\em ongoing} LHC.
Of course, simply discovering these few-TeV particles in flavor-blind channels -- even if very exciting! -- would not 
quite constitute a smoking-gun of top/Higgs compositeness which lies at the core of this 
framework.
Remarkably, we have seen above that the {\em non}-universal
contributions to the couplings of these few TeV particles -- stemming from top/Higgs compositeness --
are not far behind. Hence, precision studies of these new states can indeed 
unravel these effects.
Clearly, this 
sensitivity to $O(10)$ TeV compositeness scale is intermediate between explicit production of compositeness physics by a 100 TeV collider and indirect low energy flavor tests. 
%still indirect (cf.~that provided by a 100 TeV hadron collider) -- in this sense it is like/resembles flavor/CP tests; nonetheless it is more direct (than flavor experiments), i.e., revealing of the framework at hand.
%

%
Finally,
%
%For completeness' sake, 
%
we mention (other) {\em virtual} effects of this $\Lambda_{ \rm Higgs}$ physics
at the LHC such as on precision Higgs or top couplings measurements
or analysis of {\em continuum} top/Higgs production.
However, given $\Lambda_{ \rm Higgs } \sim O(10)$ TeV, even the high-luminosity LHC will not be sensitive to the effects in these searches.
The point is that such probes 
lack the resonance-enhancement\footnote{In fact, these states are quite {\em narrow}. For example, with the assumptions
made above and for $\Lambda_{ \rm Higgs } \sim 15$ TeV, we estimate that 
$\Gamma / M$ for 
KK $Z$ is
$O( 0.1 \% )$.
%
%\ka{Please check that this is just the ratio for SM $Z$ $\times \left( g_{  \rm EW } / g^{ \rm gauge }_{ \star } \right)^2$}
%
} that the above {\em lighter} spin-1 studies afford:
again, both these effects do have a (common) $\left( \hbox{few TeV}
%
%\Lambda_{ \rm IR } 
%
/ \Lambda_{ \rm Higgs } \right)^2$ suppression.

\subsection{Spin-2 composite}

The (total) cross-sections (again, from gluon fusion) are $\sim $ 450 (30) fb for $\Lambda_{ \rm IR } = 2 ( 3)$ TeV for $g^{ \rm grav }_{ \star \; \rm IR } = 3$ and $g_{ \star }^{ \rm QCD } = 3$.
Just like for dilaton mentioned above, 
decays are dominated by two SM gauge bosons, unless $m_{
\varphi 
%
%\rm dilaton
%
} < \frac{1}{2} M_{\rm KK grav}$,
%dilaton is lighter than $1/2$ spin-2 mass, 
in which case, decay to the dilatons dominates (due to stronger coupling).
Furthermore, for the case of universal $g^{ \rm gauge }_{ \star \; \rm IR }$, we get coupling of spin-2 to two SM gauge bosons $\propto$ corresponding
(SM gauge coupling)$^2$. Thus, 
(neglecting decays to dilaton, for example, assuming $m_{
\varphi 
%
%\rm dilaton
%
} > \frac{1}{2} M_{\rm KK grav}$)
%dilaton being heavierthan $1/2$ spin-2 mass) 
BR's to $\gamma \gamma$, $ZZ$, $WW$,  and $gg$ are $\approx$ 0.1 $\%$, $\lesssim$ 4$\%$, 4$\%$ and 90$\%$, respectively (like for radion).
%
%--- just like for dilaton, 
%
It is also clear that 
current bounds on cross-section from resonant di-photon search are
satisfied for above choice of parameters, since there is not much difference between spin-0 and spin-2 here in so far as {\em experimental} bounds are concerned.

\noindent {\bf Significance of spin 2}

Even though the final state for composite/KK graviton might be similar to dilaton (i.e., two SM gauge bosons), obviously,
spin-2 vs.~spin-0 can be distinguished using angular distributions.
In fact, 
as already mentioned earlier, a random spin-2 has three different angular amplitudes \cite{Panico:2016ary} vs.~KK graviton
having only one (i.e., coupling to $T_{ \mu \nu }$ only), hence providing disambiguation between
generic strong dynamics and extra-dimensional frameworks (i.e., dual to a {\em special} structure of strong dynamics).
Finally, it is interesting that 
mere discovery of spin-2 implies that there is an infinite tower of heavier states (whether composite or KK) because the theory of (massive) spin-2 
is
%
%has to be 
%
{\em non}-renormalizable (vs.~spin-0 or 1), thus guaranteeing more and rich discoveries in the future!

\section{Model with {\em two} intermediate branes}

\label{4brane}

Our work opens up other possibilities also: 
most significantly, we can have the gauge brane split (at $\Lambda_{ \rm meson }$) from gravity 
($\Lambda_{ \rm glueball }$) as in Fig.~\ref{fig:Multibrane}.
In this case, 
KK graviton/radion will be the lightest;
in particular, radion can be lighter than KK graviton, as seen from 
Eq.~(\ref{radion_mass})\footnote{In fact, (very) recently \cite{Georgi:2016xhm} studied a 4D model (with new -- pure glue -- strong dynamics) 
which is sort of dual of the above gauge-gravity split case (with 
%
%750 GeV being 
%
the lightest scalar glueball being roughly the radion/dilaton).}.
So, we have (parametrically speaking) $m_{ \varphi } \lesssim \Lambda_{ \rm glueball } \ll \Lambda_{ \rm meson } \ll \Lambda_{ \rm Higgs }$.
Note that the 
lightest states here are SM neutral so that 
in order to produce them, they have to couple to SM particles via exchange of heavier (for example, spin-1) composites.
%
%Nonetheless
%
In spite of the resulting 
weak couplings of 
these gravity sector states,
it is possible that they give first discoveries in parts of parameter space, especially if they sufficiently light.
Also,
stabilization of the inter-brane separations (in this case, we have {\em three} of them)
can be done via a generalization of what was done for the model with one intermediate brane above.

In more detail, the couplings of
KK graviton and radion to SM {\em gauge} bosons will be suppressed by
$\left( \Lambda_{ \rm meson } /  \Lambda_{ \rm glueball } \right)^4$ in this model, similarly to the case of
their couplings to top/Higgs in the model of Fig.~\ref{fig:3brane} studied here.
Essentially, we perform the replacements 
$T_{ \mu \nu }^{ (t/H) } \rightarrow T_{ \mu \nu }^{ \rm (gauge) }$ in Eq.~(\ref{graviton_topHiggs}) and 
Higgs kinetic term $\rightarrow F_{ \mu \nu } F^{ \mu \nu }$ in Eq.~(\ref{radion_topHiggs}), along with 
$\Lambda_{ \rm Higgs } \rightarrow \Lambda_{ \rm meson }$ in  both equations.
On the other hand, couplings of dilaton/spin-2 to {\em top/Higgs} and those of spin-1 to {\em all} SM matter remain the same.
Here, we simply summarize all these couplings in table \ref{table_coupling_4brane} (cf.~table \ref{table_coupling_3brane}).

%%%%%%%%%%%%%%%%%%%%%%%%%%%%%%%%%%%%%%%
% TABLE
%%%%%%%%%%%%%%%%%%%%%%%%%%%%%%%%%%%%%%%
\begin{table}[tbp]
\begin{adjustwidth}{-1.3cm}{}

\begin{tabular}{ | c || c | c | }
\hline 

\backslashbox[45mm]{Resonance}{Type of coupling}
&\makebox[8em]{Higgs compositeness-sensitive}&\makebox[8em]{flavor-blind}\\ 
\hline\hline

dilaton & $\frac{ g^{ \rm grav }_{ \star } }{ \Lambda_{ \rm glueball } }
\left( \frac{ \Lambda_{ \rm glueball } }{ \Lambda_{ \rm Higgs } } \right)^4 
\Big[ m_t \bar{t} t + \left( \partial H \right)^2 \Big]$
& $ \frac{ g^{ \rm grav }_{ \star } }{ \Lambda_{ \rm glueball } } \left( \frac{ g_{ \rm SM } }{ g^{ \rm gauge }_{ \star } } \right)^2 
\left( \frac{ \Lambda_{ \rm  glueball } }{ \Lambda_{ \rm meson } } \right)^4
F^{ \mu \nu } F_{ \mu \nu }$ 
%
%(gauge bosons) 
%
\tabularnewline \hline
KK $Z$ & $g^{ \rm gauge }_{ \star } \left( \frac{ \Lambda_{ \rm glueball } }{ \Lambda_{ \rm Higgs } } \right)^2 \left( \bar{t} \gamma^{ \mu } t
+ H^{\dagger } \partial^{ \mu } H 
\right)$ & $\frac{ g_{ \rm EW }^2 }{ g^{ \rm gauge }_{ \star } } \left( \bar{q} \gamma^{ \mu } q + \bar{l} \gamma^{ \mu } l \right)$ 
({\em all} generations) 
\tabularnewline \hline
KK gluon & $g^{ \rm gauge }_{ \star } \left( \frac{ \Lambda_{ \rm glueball } }{ \Lambda_{ \rm Higgs } } \right)^2 \bar{t} \gamma^{ \mu } t$
& $\frac{ g_{ \rm QCD }^2 }{ g^{ \rm gauge }_{ \star } } \bar{q} \gamma^{ \mu } q$ 
({\em all} generations) 
\tabularnewline \hline
KK graviton & $\frac{ g^{ \rm grav }_{ \star } }{ \Lambda_{ \rm glueball } } \left( \frac{ \Lambda_{ \rm glueball } }{ \Lambda_{ \rm Higgs } } \right)^4  T_{ \mu \nu }^{(t/H)}$
& $\frac{ g^{ \rm grav }_{ \star } }{ \Lambda_{ \rm glueball } } \left( \frac{ g_{ \rm SM } }{ g^{ \rm gauge }_{ \star } } \right)^2 
\left( \frac{ \Lambda_{ \rm  glueball } }{ \Lambda_{ \rm meson } } \right)^4
T_{ \mu \nu }^{\rm (gauge)}$
\tabularnewline
\hline
\end{tabular}

\caption{Summary of universal and non-universal couplings of various composites in the model with two intermediate branes.}
\label{table_coupling_4brane}
\end{adjustwidth}
\end{table}

A sample point is as follows: $\Lambda_{ \rm meson } = 2$ TeV,
%
%(in order to satisfy bound on spin-1), 
%
$m_{ \varphi} =
%
% \sim 
%
\Lambda_{ \rm glueball } = 1$ TeV, $g^{ \rm QCD }_{ \star } = 3$ and $g^{ \rm grav }_{ \star \; \rm IR } \sim 3$ gives (total) cross-section of
$\sim 100$ fb and $\sim 30$ fb, respectively, for KK graviton and dilaton (former being larger mostly due to multiple polarizations).
The decay BR's are similar to the model with one intermediate brane case.
Note that 
gauge KK/spin-1 composite cross-section at this point are comparable to/larger than
that of graviton/dilaton; in fact, the gauge KK would be strongly constrained (if not ruled out), {\em assuming} decays directly to SM particles
(as discussed above).
However, 
the spin-1 states can 
decay directly
%
%first 
%
into non-SM particles such that they are effectively ``hidden'' from SM pair-resonance searches such as dileptons or dijets.
For example, light $A_5$'s (dual to PNGB's) can provide such channels.\footnote{For a recent application of this idea in the context of the 750 GeV
diphoton excess at the LHC, see 
\cite{Carmona:2016jhr}.}
In this way, KK graviton/dilaton 
can actually be the most visible channel.
Table \ref{table_radion_KK gravity} in the introduction had already 
displayed this interesting possibility.

Based on the discussion in section \ref{subsubsec:spin-2/KK graviton} of KK graviton contributions to precision tests, it is clear that the only relevant constraint on the KK graviton mass scale, i.e., 
$\Lambda_{ \rm glueball }$, in this model comes from direct LHC searches;
in particular, using the cross-sections given above and bounds given earlier, we see that $\Lambda_{ \rm glueball }$ is then allowed to 
be as low as $\sim 1$ TeV (or even smaller).

\section{Conclusions/Outlook}
\label{sec:conclusion}

The LHC Run 1 complemented by electroweak/flavor/CP precision tests have thus far seen no deviations from the SM. In light of this we must conclude that the principle of Naturalness, that predicts new physics below the TeV scale, is either (i) at the cusp of discovery at the LHC, (ii) playing itself out in some exceptional dynamics (such as Twin Higgs theory \cite{Chacko:2005pe}) that evades our standard experimental probes, or (iii) that the principle is compromised in some way.  Our efforts must be directed at all these options. 
Higgs compositeness (AdS/CFT dual to warped extra-dimensions) within the LHC reach remains a strongly motivated possibility for (i), but requires some new refinement of the warped GIM mechanism.
This paper is directed instead to the option (iii) in the same, broad framework.
%
% of warped extra-dimensions/composite Higgs. 
%
Indeed, it is noteworthy that 
%
%the core comprehensive naturalness mechanisms of supersymmetry and warped dimensions/compositeness 
%
the {\em minimal} incarnation of this 
%
%framework
%
paradigm (see Fig.~\ref{fig:OriginalRS}) can readily and elegantly fit the experimental facts if we take the related new physics to live at $\sim O(10)$ TeV, solving the ``big hierarchy problem'' between the electroweak and Planck scales, but leaving unexplained a ``little hierarchy problem''. It is not the modest associated fine-tuning that disturbs us here but the fact that the solution to the hierarchy problem would then lie out of LHC reach!

We have shown that a simple {\em extension} of the 
above model 
%
%minimal warped extra-dimensional model AdS/CFT dual to Higgs compositeness 
%
can also readily fit all the experimental facts if the physics of naturalness is deferred until $\sim O(10)$ TeV. 
Namely, 
%
%we showed that 
%
when different fields propagate different amounts into the IR of the extra dimension (see Fig.~\ref{fig:Multibrane}), there can naturally be lighter TeV-scale ``vestiges'' of the heavy naturalness physics within LHC reach, in the form of new spin-0,1,2 resonances, identified as KK excitations of the extra dimension or composites in the dual mechanism of vectorlike confinement. Although they would constitute a rich new physics close at hand, they escape the strong constraints from flavor/CP tests by virtue of their largely flavor-blind, gauge-mediated couplings to the standard model.  We have described several striking features of their phenomenology in the 5D Randall-Sundrum framework and its AdS/CFT dual. 
In particular, search channels such as dileptons, dijets and diphotons for the $\sim$ TeV-mass resonances 
acquire significance in this framework, cf.~decays being dominated by top/Higgs in the minimal model of Fig.~\ref{fig:OriginalRS}.

But flavor-blindness, however rich the physics, also suggests blindness to the solution to the hierarchy problem. Fortunately, we saw there are small deviations from flavor-blindness in resonance decays into top/Higgs rich final states. These processes thereby give a resonance-enhanced ``preview'' of Higgs compositeness at the LHC, even though the Higgs compositeness scale and its ultimate resolution of the hierarchy problem is out of LHC reach! This provides a pathway in which LHC discoveries might set the stage for even higher energy explorations. A roughly comparable analogy within the supersymmetric paradigm is (mini-) Split SUSY 
\cite{Arvanitaki:2012ps}, in which the stops most directly relevant to the hierarchy problem lie above LHC reach (helping to explain the larger-than-expected Higgs boson mass) while spin-1/2 super-partners are significantly lighter. Seeing the lighter super-partners at the LHC with their SUSY-specific quantum numbers would also give a ``preview'' of the supersymmetric solution to the hierarchy problem, which could be fully confirmed by going to higher energy colliders.

In future work, it will be interesting to study in more detail the LHC signals for the (approximately) flavor-blind $\sim$ TeV mass 
resonances which were outlined here, including what we can learn about the physics of 
top/Higgs compositeness at $\sim O(10)$ TeV from their precision analysis. In addition, we plan to initiate investigation of more direct signals of the latter physics which might be possible at a future 100 TeV hadron collider.

We are now at the beginning of LHC Run 2, and it is essential that theory lays out the most plausible and powerful mechanisms within reach. In the language of 4D strong dynamics we have shown that vectorlike confinement arising in the IR of Higgs compositeness is such a plausible form of new physics, already exciting at the LHC and able to pave the way for an even more ambitious program of discovery at future higher-energy colliders.

\section*{Acknowledgements}

%
%\begin{eqnarray}
%\end{eqnarray}
%

%

We would like to thank Zackaria Chacko for discussions.
This work was supported in part by NSF Grant No.~PHY-1315155 and the Maryland Center for Fundamental Physics. SH was also supported in part by a fellowship from The Kwanjeong Educational Foundation.

\appendix

\section{Details of choice of parameters}

\label{choice_detail}

\subsection{Matching at the intermediate/Higgs brane}

%
%-- \ka{Please check formulae in these 2 sections!}
%

We assume the same 5D Planck scale ($M_5$) throughout the bulk.
However, in short, the bulk cosmological constant (CC) -- and hence AdS curvature {\em scale} ($k$) -- will be 
different in the matter/Higgs and gauge/gravity (only) bulks
due to presence of (tension on) the intermediate/Higgs brane.
In more detail, we define
\bea
\hbox{CC}_{ \rm UV \; (IR ) } & \equiv & 24 M_5^3 k_{ \rm UV \; (IR) }^2,  \nonumber \\
k _{ \rm UV \; (IR) } & \equiv & \frac{1}{ R^{ \rm AdS }_{ \rm UV \; (IR) } }
\eea
where ``UV'' and ``IR'' denote the
bulks on the two sides of the Higgs brane and 
$R^{ \rm AdS }$ is the AdS curvature radius.
%
%matching at 
%
Solving Einstein's equations across the 
the Higgs brane (with tension, $T_{ \rm Higgs }$) gives \cite{Kogan:2001qx}:
\bea
T_{ \rm Higgs } & = & 12 M_5^3 \left( k_{ \rm  IR } - k_{ \rm UV } \right)
\label{T_Higgs}
\eea
%
%i.e., the curvature scales on the 2 sides are different...
%
Since we require $T_{ \rm Higgs } > 0$ in order to avoid a branon (brane-bending degrees of freedom, denoted by $Y$) ghost
\cite{Sundrum:1998sj}, we
see that 
\bea
k_{ \rm UV } & < & k_{ \rm IR }
\eea
i.e., curvature scale {\em increases} in the IR.
Let us consider in the following how this new feature modifies 
the usual choice of parameters.

\subsection{Implications of above matching} 

Consider the gravity sector of the model first. Clearly, we then have two different $g_{ \star }^{ \rm grav }$'s on the two sides of
the Higgs brane:
\bea
\frac{ g^{ \rm grav}_{\star \; \rm IR } }{ g^{ \rm grav}_{\star  \; \rm UV } } & = & \sqrt{ \frac{ k_{ \rm IR }^3 }{ k_{ \rm UV }^3 } }
\eea
As usual,  
we have bulk gravity becoming strongly coupled at \cite{Chacko:1999hg}
\bea
\Lambda^{ \rm grav }_{ \rm strong } & \sim & k \left( \frac{ 16 \pi^2 }{ g^{ \rm grav \; 2 }_{ \star } } \right)^{ 1/3 }
\eea
Suppose we would like to have at least $N^{ \rm min }_{ \rm KK }$ number of weakly-coupled KK modes
(i.e., that much gap between 5D cut-off and curvature scale as our control parameter).
Then
we must have
\bea
g^{ \rm grav }_{ \star } & \lesssim & \sqrt{ \frac{ 16 \pi^2 }{ N^{ \rm min \ 3 }_{ \rm KK } } } 
\eea
from the condition that $\Lambda^{ \rm grav }_{ \rm strong } \gtrsim N^{ \rm min }_{ \rm KK } k$.
Note that 
this is required in each of the two bulks, i.e., for both $g_{ \star \; \rm IR, \; UV }^{ \rm grav }$.
Of course, in order to avoid large hierarchies amongst fundamental/5D parameters
(for example, between $k$ and $M_5$), we would
also impose that $g^{ \rm grav }_{ \star }$ is {\em not} $\ll 1$.

Moving onto {\em gauge} sector, we similarly have
\bea
\frac{ g^{ \rm gauge}_{\star \; \rm IR } }{ g^{ \rm gauge}_{\star \; \rm UV } } & = & \sqrt{ \frac{ k_{ \rm IR } }{ k_{ \rm UV } } }
\eea
The 
strong coupling scale is given by :
\bea
\Lambda^{ \rm gauge }_{ \rm strong } & \sim & k \frac{ 16 \pi^2 }{ N_{ \rm SM } \; 3 \; g^{ \rm gauge \; 2 }_{ \star } } 
\eea
where $N_{ \rm SM }$ denotes size of the SM gauge group (take it here to be 3 for color) and factor of 3 in denominator
above (i.e., enhancement of loop expansion parameter) comes from counting helicities of spin-1 field.
So, the associated request (i.e, imposing $\Lambda^{ \rm gauge }_{ \rm strong } \gtrsim N^{ \rm min }_{ \rm KK } k$) is
\bea
g^{ \rm gauge }_{ \star } & \lesssim & \frac{ 4 \pi }{ 3 \sqrt{ \; N^{ \rm min }_{ \rm KK } } }
\eea
for each of the two bulks.

On the other hand, fitting to
the observed/SM gauge coupling gives {\em lower} limits on $g^{ \rm gauge }_{ \star }$ as follows
(note that there is no analog of Landau pole for gravity, hence no lower limit on $g^{ \rm grav}_{ \star }$ on this count).
Consider the running of the SM gauge couplings 
from the UV cut-off to the
IR shown in Eq.~(\ref{gauge_match}).
Plugging in the low-energy values of $g_{ \rm SM }$ and $b_{ \rm SM }$ into Eq.~(\ref{gauge_match}), we find
(assuming $\Lambda_{\rm Higgs} \sim 10 \; {\rm TeV}$ and $\Lambda_{\rm IR} \sim$ few $\rm TeV$)
\bea
g_{ \star \; \rm UV }^{ \rm gauge } & \gtrsim & 3 
\label{Landau}
\eea
from the requirement that $ 1 / g_{ \rm UV }^2  > 0$, i.e., Landau poles for SM gauge couplings are at/above 
$\sim 10^{ 15 }$ GeV.
However, $g^{ \rm gauge }_{ \star \; \rm IR }$ mostly unconstrained, since it contributes over a (much) smaller hierarchy.

%
%However, 
%
Finally, there is another requirement that the strong coupling scale of the
$Y$ self-interactions be (at least modestly) above the curvature scale, i.e.,
\bea
{\cal L} & \ni & \left( \partial_{ \mu } Y \right)^2 + \frac{ \left( \partial_{ \mu } Y \right)^4 }{ T_{ \rm Higgs } }
\eea
results in
\bea
\Lambda^{ \rm branon }_{\rm strong} & \sim & \left( 16 \pi^2 T_{ \rm Higgs } \right)^{ 1/4 }
\eea
So we need [as usual, imposing $\Lambda^{ \rm branon }_{ \rm strong } \gtrsim N^{ \rm min }_{ \rm KK } k $ and using
Eq.~(\ref{T_Higgs})]
\bea
g^{ \rm grav \; IR }_{ \star } & \lesssim & \Big[ \frac{ 192 \pi^2 }{ N^{ \rm min \; 4 }_{ \rm KK } } \left( 1 - \frac{ k_{ \rm UV } }{ k_{ \rm IR } }
\right)  \Big]^{ 1/2 } 
\eea

We can check that the following choices of couplings {\em barely} satisfy {\em all} the above needs (including giving observable LHC signals):
\bea
g^{ \rm grav}_{ \star \; \rm UV  } < g^{ \rm grav }_{ \star \; \rm IR} \lesssim 3 & \hbox{and} & g^{ \rm gauge }_{ \star \; \rm UV } \sim 3; \;\;
g^{ \rm gauge \; 2 }_{ \star \; \rm IR} \approx g^{ \rm gauge \; 2 }_{ \star \; \rm UV } \Big[ 1 + O (0.1) \Big]  
%
%\sim 4-5
%
\eea
for a 
%
%reasonable 
%
{\em minimal}
request of 
\bea
N^{ \rm min }_{ \rm KK } & \sim & 2
\eea
(and corresponding to $k_{ \rm IR } / k_{ \rm UV } \approx 
1 + O (0.1)$).
%
%$\sim 1/2-1/3$.
%

Note that $g^{ \rm gauge }_{ \star \; \rm IR }$ and $g^{ \rm gauge }_{ \star \;  \rm UV }$ are ``forced'' to be close to each other,
due to a {\em combination} of perturbativity (upper bound on $g^{ \rm gauge }_{ \star \; \rm UV }$) and Landau pole (lower bound) 
constraints. 
One possibility to relieve this tension is to reduce the UV-IR hierarchy, for example, lower the UV scale to the flavor scale of 
$\sim 10^5$ TeV \cite{Davoudiasl:2008hx}, while keeping IR scale $\sim$ TeV: from Eq.~(\ref{gauge_match}), we see that $g^{ \rm gauge }_{ \star \; \rm UV } 
\gtrsim 2$ is then allowed (keeping both $g^{ \rm gauge }_{ \star }$'s at/below $\sim 3$ for perturbativity).

\section{Two Dilaton system}

\label{2-dilaton}

Here we discuss the CFT dual of stabilization of the model with one intermediate brane studied in the main text.
In short, as usual, we start with a CFT at a UV cut-off $\Lambda_{ \rm UV }$.
This CFT confines, i.e., scale invariance is broken, at $\Lambda_{ \rm int }$, which is to be identified with $\Lambda_{ \rm Higgs}$, i.e., 
scale of the Higgs brane in the specific model, but here we would like to keep the notation more general.
As already mentioned, this scale can be 
``parametrized'' by VEV of dilaton/radion field [denoted by $\Phi_{ \rm int }$ of mass dimension $+1$, fluctuations around which correspond to the physical dilaton ($\varphi_{ \rm int }$)], i.e.,
\bea
\Phi_{ \rm int } & \sim  & \Lambda_{ \rm int } + a g^{ \rm grav }_{ \star \; \rm UV } \varphi_{ \rm int }. 
\eea
The departure from the standard (i.e., minimal model of Fig.~\ref{fig:OriginalRS}) script involves the resulting (daughter) theory (i.e., below
$\Lambda_{ \rm int }$) flowing to a new fixed point.
This ``IR'' CFT then confines at an even lower scale $\Lambda_{ \rm IR }$, corresponding to the VEV of {\em another} field, $\Phi_{ \rm IR }$
(associated with a second dilaton, $\varphi_{ \rm IR }$).

In more detail, in order to stabilize the two inter-brane separations (dual to determining the various mass scale hierarchies), we perturb the CFT by adding a {\em single} scalar operator %of scaling dimension
(dual to the GW field) in the UV:
\bea
{\cal L} \left( \Lambda_{ \rm UV } \right) & \ni & {\cal L}_{ \rm CFT \; UV } + \lambda \; \Lambda_{ \rm UV }^{ \epsilon_{ \rm UV} } 
\; {\cal O}^{ \rm UV }_{ \rm GW } 
\label{LCFT_UV}
\eea
where scaling {\em and} naive/engineering dimension of ${\cal O}_{ \rm GW }^{ \rm UV }$ is 
$( 4 - \epsilon_{ \rm UV} )$ (i.e., $\lambda$ above is dimension{\em less}).
As usual, we assume that there is only one scalar operator with scaling dimension close to 4, rest of them
being irrelevant (hence being dropped from the Lagrangian).
We flow to $\Lambda_{ \rm int }$ (as usual, promoting appropriately $\Lambda$'s to $\Phi$'s throughout):
\bea
{\cal L} \left( \Lambda_{ \rm int } \right) & \ni & {\cal L}_{ \rm CFT \; IR } + \left( \partial_{ \mu } \Phi_{ \rm int } \right)^2 + \lambda^{ \prime } \Phi_{ \rm int }^4 + 
d_1 \lambda \; 
\Phi_{ \rm int }^4 \left( \frac{  \Phi_{ \rm int }  }{ \Lambda_{ \rm UV } } \right)^{ - \epsilon_{ \rm UV } } + \nonumber \\
& & \Big[ d_2 \lambda \left( \frac{  \Phi_{ \rm int } 
%
%\Phi_{ \rm int }  
%
}{ \Lambda_{ \rm UV } } \right)^{ - \epsilon_{ \rm UV } }  + 
\tilde{ \lambda } 
%
%\left(  \Lambda_{ \rm int } 
%
%\Phi_{ \rm int } 
%
%\right) 
%
\Big] \Phi_{ \rm int }^{ \epsilon_{ \rm IR } } {\cal O}_{ \rm GW }^{ \rm IR } 
%+  \mathcal{O} \left( \lambda^2 \left( \frac{  \Phi_{ \rm int }  }{ \Lambda_{ \rm UV } } \right)^{ - 2 \epsilon_{ \rm UV } } \right)
\label{LCFT_int}
\eea
where $d_{ 1, \; 2 }$ are O(1) factors.

Let us elaborate on the various terms above.
The first three terms above (in first line) are as discussed earlier (i.e., for the usual minimal model).
Whereas, the 
first {\em new} term (in second line above) comes from using the interpolation:
\bea
{\cal O}_{ \rm GW }^{ \rm UV } \left( \Lambda_{ \rm int } \right ) & \sim &{\cal O}_{ \rm GW }^{ \rm IR } 
\Phi_{ \rm int }^{ \epsilon_{ \rm IR } - \epsilon_{ \rm UV } } +...
\eea
in the RG evolved {\em explicit} conformal symmetry breaking term in Eq.~(\ref{LCFT_UV}).
Here, (with obvious choice of notation) ${\cal O}_{ \rm GW }^{ \rm IR }$ is an operator of the {\em IR} 
CFT of scaling dimension $( 4 - \epsilon_{ \rm IR } )$: again, we assume that there
exists only one such operator.
On the other hand, the second term in second line of Eq.~(\ref{LCFT_int}) arises from {\em spontaneous} conformal symmetry breaking
at scale $\Lambda_{ \rm int }$, i.e.,
even if $O_{ \rm GW }^{ \rm UV }$ were ``absent''.
Given above assumption about scaling dimensions of scalar operators of the {\em IR} CFT, it is clear that
both terms in second line above must involve the {\em same} operator (as the leading term), i.e., coupling of $\Phi_{\rm int}$ to {\em other} scalar operators of the IR CFT will be irrelevant.
%
%Note that (in general) we will also have terms coupling $\Phi_{ \rm int }$ to ${\cal O}_{ \rm GW} ^{ \rm IR }$ (and similarly other scalar operators); however, such effects are irrelevant (again given above assumption about IR CFT).

Finally, i.e., RG flowing to the far IR scale of $\Lambda_{ \rm IR }$ and adding (for a second time) the usual term 
consistent with the (IR) conformal symmetry, we obtain the complete potential for the two scalar fields ($\Phi$'s):
\bea
{\cal L} \left( \Lambda_{ \rm IR } \right) & \ni & \left( \partial_{ \mu } \Phi_{ \rm int } \right)^2 
 + \lambda^{ \prime } \Phi_{ \rm int }^4 + 
d_1 \lambda \Phi_{ \rm int }^4 \left( \frac{  \Phi_{ \rm int }  }{ \Lambda_{ \rm UV } } \right)^{ - \epsilon_{ \rm UV } } +  %\mathcal{O} \left( \lambda^2 \left( \frac{  \Phi_{ \rm int }  }{ \Lambda_{ \rm UV } } \right)^{ - 2 \epsilon_{ \rm UV } } \right) 
\nonumber \\
& & 
\left( \partial_{ \mu } \Phi_{ \rm IR } \right)^2  + \tilde{ \lambda }^{ \prime } \Phi_{ \rm IR }^4 +  \Big[ 
d_3 \lambda 
\left( \frac{  \Phi_{ \rm int } }{ \Lambda_{ \rm UV } } \right)^{ - \epsilon_{ \rm UV } }  + 
d_4 \tilde{ \lambda } \Big] 
\left( \frac{ \Phi_{ \rm IR } }{ \Phi_{ \rm int } } \right)^{ - \epsilon_{ \rm IR } } \Phi^4_{ \rm IR } %+ \nonumber \\ & & \mathcal{O} \left( \lambda^2, \; \tilde{\lambda}^2, \; \left( \frac{  \Phi_{ \rm int } }{ \Lambda_{ \rm UV } } \right)^{ - 2 \epsilon_{ \rm UV } } , \; \left( \frac{ \Phi_{ \rm IR } }{ \Phi_{ \rm int } } \right)^{ - 2 \epsilon_{ \rm IR } } \right)
\label{V_2dilaton}
\eea

We have to minimize the above potential in order to determine the scales $\Lambda_{ \rm int }$ and $\Lambda_{ \rm IR }$
in terms of $\Lambda_{ \rm UV }$ and the scaling dimensions [we can assume that the various
$\lambda$'s are $O(1)$].
As usual, we assume $\epsilon_{ \rm UV, \; IR }$ are {\em modestly} smaller than 1.
In this case, we can proceed with the minimization in steps as follows.
At ``leading-order'' (LO), it is reasonable to assume that $\langle \Phi_{ \rm int } \rangle \sim \Lambda_{ \rm int }$ is 
mostly determined (as in the minimal two brane case) by first line of Eq.~(\ref{V_2dilaton}) (i.e., potential for 
$\Phi_{ \rm int }$ by itself) to be:
%
%\bea
%\Lambda_{ \rm int } & \sim & \Lambda_{ \rm UV } e^{ - O \left( \frac{1}{ \epsilon_{ \rm UV} } %\right) }  
%\eea
%
\bea
\Lambda_{ \rm int } & \sim & \left( -d_1 \frac{\lambda}{\lambda'} \right)^{1/\epsilon_{\rm UV}} \Lambda_{ \rm UV }   
\eea
with 
\bea
m_{ \varphi_{ \rm int } }^2 & \sim \epsilon_{ \rm UV } \Lambda_{ \rm int }^2.
\label{heavy_dilaton}
\eea
Similarly, plugging $\Phi_{ \rm int } = \Lambda_{ \rm int }$ (i.e., a fixed value) into second line of Eq.~(\ref{V_2dilaton}), i.e., effective potential for $\Phi_{ \rm IR }$, will give (again, as usual):
%
%\bea
%\Lambda_{ \rm IR} & \sim & \Lambda_{ \rm int } e^{ - O \left( \frac{1}{ \epsilon_{ \rm IR} } \right) }  
%\eea
%
\bea
\Lambda_{ \rm IR} & \sim & \left( - \frac{1}{\tilde{\lambda}^{\prime}} \left( d_4 \tilde{\lambda} - \frac{d_3}{d_1} \lambda' \right) \right)^{1/\epsilon_{\rm IR}} \Lambda_{ \rm int } 
\eea
with 
\bea
m_{ \varphi_{ \rm IR } }^2 & \sim \epsilon_{ \rm IR } \Lambda_{ \rm IR }^2. 
\label{light_dilaton}
\eea

As a (partial) consistency check of the above procedure (for obtaining the values of VEV's), we can consider the mixing (if you will, the NLO) term involving {\em both} the dilatons arising from the last two 
terms of second line of Eq.~(\ref{V_2dilaton}), where $\Phi_{ \rm int }$ can be thought of as 
fluctuations around $\Lambda_{ \rm int }$:
\bea
\Delta m^2_{ \varphi_{ \rm int } \varphi_{ \rm IR } }
& \sim & O \left(  \epsilon_{ \rm IR }, \; \epsilon_{ \rm UV } \right)  \frac{ \Lambda^3_{ \rm IR } }{ \Lambda_{ \rm int } }:
\eea
we see that this results in a mixing angle between two dilatons of $\sim \epsilon \left( \Lambda_{ \rm IR } / \Lambda_{ \rm int } \right)^3$, i.e., small enough.
As a further check, we can show that the first derivatives of the {\em full} 
potential in Eq.~(\ref{V_2dilaton}) at above values of VEV's vanish, up to 
terms suppressed by (powers of) $\Lambda_{ \rm IR } / \Lambda_{ \rm int }$, i.e., the actual VEV's are close enough to
those obtained by the above ``piece-wise'' minimization of the potential.
Hence, to a good approximation, we can ``decouple'' the two dilaton systems (as already assumed in the main text).

\end{document}